\documentclass[11pt, draftcls,onecolumn]{IEEEtran}

\usepackage{amsmath,amstext,amsfonts,amssymb,amsthm,bbold} 
\usepackage{graphicx,subfigure,color} 
\usepackage{hyperref}
\usepackage{epstopdf}
\usepackage[justification=centering]{caption}
\def \A {\mathbf{A}}
\def \a {\mathbf{a}}

\def \d {\mathbf{d}}

\def \e {\mathbf{e}}

\def \H {\mathbf{H}}
\def \I {\mathbf{I}}

\def \J {\mathbf{J}}

\def \Q {\mathbf{Q}}

\def \s {\mathbf{s}}
\def \S {\mathbf{S}}

\def \U {\mathbf{U}}
\def \u {\mathbf{u}}
\def \v {\mathbf{v}}
\def \V {\mathbf{V}}
\def \W {\mathbf{W}}

\def \x {\mathbf{x}}

\def \Y {\mathbf{Y}}
\def \y {\mathbf{y}}

\def \Ical {\mathcal{I}}

\def \Ncal {\mathcal{N}}
\def \Ocal {\mathcal{O}}

\def \Cbb {\mathbb{C}}
\def \Ebb {\mathbb{E}}
\def \Nbb {\mathbb{N}}
\def \Pbb {\mathbb{P}}
\def \Rbb {\mathbb{R}}

\def \drm {\mathrm{d}}
\def \erm {\mathrm{e}}
\def \irm {\mathrm{i}}

\def \det {\mathrm{det}}

\def \Tr {\mathrm{Tr}\,}

\DeclareMathOperator{\sinc}{sinc}
\DeclareMathOperator*{\argmin}{argmin}

\newcommand{\bs}{\boldsymbol}

\newtheorem{theorem}{Theorem}

\newtheorem{remark}{Remark}

\newtheorem{proposition}{Proposition}
\newtheorem{assum}{Assumption}

\begin{document}
%
\title{Performance analysis of spatial smoothing schemes in the context of large arrays.}
%
%
%

\author
{       Gia-Thuy Pham,
        Philippe~Loubaton,~\IEEEmembership{Fellow,~IEEE}
	and~Pascal~Vallet,~\IEEEmembership{Member,~IEEE,}
	\thanks
	{
		P. Vallet is with Laboratoire de l'Int\'egration du Mat\'eriau au Syst\`eme (CNRS, Univ. Bordeaux, IPB),
		351, Cours de la Lib\'eration 33405 Talence (France), pascal.vallet@ipb.fr
	}
	\thanks
	{
		G.T. Pham and P. Loubaton are with Laboratoire d'Informatique Gaspard Monge (CNRS, Universit\'e Paris-Est/MLV), 5 Bd. Descartes 77454 Marne-la-Vall\'ee (France),
		loubaton@univ-mlv.fr
	}
}

\maketitle

\begin{abstract}
	This paper adresses the statistical behaviour of spatial smoothing subspace DoA estimation schemes using a sensor array in the case where the number of observations $N$ is significantly smaller than the number of sensors $M$, and that the smoothing parameter $L$ is such that $M$ and
$NL$ are of the same order of magnitude. This context is modelled by an asymptotic
regime in which $NL$ and $M$ both converge towards $\infty$ at the same rate. As in recent works devoted to the study of (unsmoothed) subspace methods
in the case where $M$ and $N$ are of the same order of magnitude, it is shown that
it is still possible to derive improved DoA estimators termed as Generalized-MUSIC with spatial smoothing (G-MUSIC SS). The key ingredient of this work is a technical result showing that the largest singular values and corresponding singular vectors of low rank deterministic perturbation of certain Gaussian block-Hankel large random matrices behave as if the entries of the latter random matrices were independent identically distributed. This allows to conclude that when the number of sources and their DoA do not scale with
$M,N,L$, a situation modelling widely spaced DoA scenarios, then both traditional and Generalized spatial smoothing subspace methods provide consistent DoA estimators whose convergence speed is faster than $\frac{1}{M}$. The case of DoA that are spaced of the order of a beamwidth, which models closely spaced sources, is also
considered. It is shown that the convergence speed of G-MUSIC SS estimates is unchanged, but
that it is no longer the case for MUSIC SS ones.
\end{abstract}


\section{Introduction}
\label{section:introduction}
The statistical analysis of subspace DoA estimation methods using an array of sensors
is a topic that has received a lot of attention since the seventies. Most of the works
were devoted to the case where the number of available samples $N$ of the observed signal is much larger than the number of sensors $M$ of the array (see e.g. \cite{stoica1989music} and the references therein). More recently, the case where $M$ and $N$ are large and of the same order of magnitude was addressed for the first time in \cite{mestre2008improved} using large random matrix theory. \cite{mestre2008improved}
was followed by various works such as \cite{johnson-abramowitch-mestre-2008}, \cite{vallet2012improved}, \cite{hachem2012large}, \cite{hachem2012subspace}. The number of
observations may also be much smaller than the number of sensors. In this context, it is well established
that spatial smoothing schemes, originally developped to address coherent sources (\cite{evans-johnson-sun-1982}, \cite{shan-wax-kailath-1985}, \cite{pillai-kwon-1989}), can be used to artificially
increase the number of snapshots (see e.g. \cite{stoica1989music} and the references therein,
see also the recent related contributions \cite{thakre-haardt-giridhar-2010-SPL},  \cite{thakre-haardt-giridhar-2010} devoted to the case where $N=1$). Spatial smoothing consists in considering $L < M$ overlapping arrays with $M-L+1$ sensors, and allows to generate artificially $NL$ snapshots
observed on a virtual array of $M-L+1$ sensors. The corresponding $(M-L+1) \times NL$
matrix, denoted ${\bf Y}_N^{(L)}$, collecting the observations is the sum of a low rank component generated by
$(M-L+1)$-dimensional steering vectors with a noise matrix having a block-Hankel structure.
Subspace methods can still be developed, but the statistical analysis of the
corresponding DoA estimators was addressed in the standard regime where $M-L+1$ remains
fixed while $NL$ converges towards $\infty$. This context is not the most relevant
when $M$ is large because $L$ must be chosen in such a way that the number of virtual
sensors $M-L+1$ be small enough w.r.t. $NL$, thus limiting the statistical performance of the estimates.
In this paper, we study the statistical performance of spatial smoothing subspace DoA
estimators in asymptotic regimes where $M-L+1$ and $NL$ both converge towards $\infty$ at the same rate, where $\frac{L}{M} \rightarrow 0$ in order to not affect the aperture of the
virtual array, and where the number of sources $K$ does not scale with $M,N,L$. For this,
it is necessary to evaluate the behaviour of the $K$ largest eigenvalues and corresponding eigenvectors of the empirical covariance matrix $\frac{{\bf Y}_N^{(L)} {\bf Y}_N^{(L)*}}{NL}$. To address this issue,
we prove that the above eigenvalues and eigenvectors have the same asymptotic behaviour as if
the noise contribution ${\bf V}_N^{(L)}$ to matrix ${\bf Y}_N^{(L)}$, a block-Hankel random matrix, was a Gaussian random matrix with independent identically distributed. To establish this result, we rely on the recent result \cite{loubaton-bloc-hankel} addressing the behaviour of the singular values of large block-Hankel random matrices built from i.i.d. Gaussian sequences. \cite{loubaton-bloc-hankel} implies that the empirical eigenvalue distribution of matrix $\frac{{\bf V}_N^{(L)} {\bf V}_N^{(L)*}}{NL}$ converges towards the Marcenko-Pastur distribution, and that its eigenvalues are almost surely located
in the neigborhood of the support of the above distribution. This allows to generalize the
results of \cite{benaych2012singular} to our random matrix model, and to characterize
the behaviour of the largest eigenvalues and eigenvectors of $\frac{{\bf Y}_N^{(L)} {\bf Y}_N^{(L)*}}{NL}$. We deduce from this improved subspace estimators, called DoA G-MUSIC SS (spatial smoothing) estimators, which are similar
to those of \cite{vallet2012improved} and \cite{hachem2012subspace}. We deduce from the results of \cite{vallet-loubaton-mestre-2014} that when the DoAs do not scale with $M,N,L$, i.e. if the DoAs are widely spaced compared to aperture array, then both G-MUSIC SS and traditional MUSIC SS estimators
are consistent and converge at a rate faster than $\frac{1}{M}$. Moreover, when the DoAs are spaced of the order
of $\frac{1}{M}$, the behaviour of G-MUSIC SS estimates remains unchanged, but
the convergence rate of traditional subspace estimates is lower.

This paper is organized as follows. In section \ref{sec:formulation}, we precise the
signal models, the underlying assumptions, and formulate our main results.
In section \ref{sec:technical-results}, we prove that the largest singular values and corresponding singular vectors of low rank deterministic perturbation of certain Gaussian block-Hankel large random matrices behave as if the entries of the latter random matrices were independent identically distributed.
In section \ref{sec:utilisation}, we apply the results of section \ref{sec:technical-results}
to matrix ${\bf Y}_N^{(L)}$, and follow \cite{hachem2012subspace}
in order to propose a G-MUSIC algorithm to the spatial smoothing  context of
this paper. The consistency and the convergence speed of the G-MUSIC SS estimates and of
the traditional MUSIC SS estimates are then deduced from the results of \cite{vallet-loubaton-mestre-2014}. Finally, section \ref{sec:simulation} present numerical experiments sustaining our theoretical results.

\textit{Notations :}
For a complex matrix $\A$, we denote by $\A^T, \A^*$ its transpose and its conjugate transpose, and by $\Tr(\A)$ and $\|\A\|$ its trace and spectral norm.
The identity matrix will be $\I$ and $\e_n$ will refer to a vector having all its components equal to $0$ except the $n$-th equals to $1$.
For a sequence of random variables $(X_n)_{n \in \Nbb}$ and a random variable $X$, we write
\begin{align}
	X_n \xrightarrow[n\to\infty]{a.s.} X
	\notag
\end{align}
when $X_n$ converges almost surely towards $X$. Finally, $X_n = o_{\Pbb}(1)$ will stand for the convergence of $X_n$ to $0$ in probability, and
$X_n = \Ocal_{\Pbb}(1)$ will stand for tightness (boundedness in probability).

		\section{Problem formulation and main results.}
\label{sec:formulation}
\subsection{Problem formulation.}

We assume that $K$ narrow-band and far-field source signals are impinging on a uniform linear array of $M$ sensors, with $K < M$. In this context, the $M$--dimensional received signal $(\y_n)_{n \geq 1}$ can be written as
\begin{align}
	\y_n = \A_M \s_n + \v_n,
	\notag
\end{align}
where
\begin{itemize}
	\item $\A_M = [\a_M(\theta_1),\ldots,\a_M(\theta_K)]$ is the $M \times K$ matrix of $M$--dimensionals steering vectors $\a_M(\theta_1),\ldots,\a_M(\theta_K)$, with
	$\theta_1,\ldots,\theta_K$ the source signals DoA, and $\a_M(\theta)= \frac{1}{\sqrt{M}}[1,\ldots,\erm^{\irm (M-1) \theta}]^T$ ;
	\item $\s_n \in \Cbb^K$ contains the source signals received at time $n$, considered as unknown deterministic ;
	\item $(\v_n)_{n \geq 1}$ is a temporally and spatially white complex Gaussian noise with spatial covariance $\Ebb[\v_n\v_n^*]=\sigma^2 \I$.
\end{itemize}
The received signal is observed between time $1$ and time $N$, and we collect the available observations in
the $M \times N$ matrix $\Y_N$ defined
\begin{align}
	\Y_N = [\y_1,\ldots,\y_N] = \A_M \S_N + \V_N,
	\label{eq:model_signal_N}
\end{align}
with $\S_N=[\s_1,\ldots,\s_N]$ and $\V_N = [\v_1,\ldots,\v_N]$. We assume that $\mathrm{Rank}({\bf S}_N) = K$
for each $M,N$ greater than $K$. The DoA estimation problem consists in estimating the $K$ DoA $\theta_1,\ldots,\theta_K$
from the matrix of samples $\Y_N$.

When the number of observations $N$ is much less than the number of sensors $M$, the standard subspace method fails. In this case, it is standard to use spatial smoothing schemes in order to artificially increase the number of observations. In particular,
it is well established that spatial smoothing schemes allow to use subspace methods even in the single snapshot case, i.e. when $N=1$ (see e.g. \cite{stoica1989music} and the references therein). If $L < M$,
spatial smoothing consists in considering $L$ overlapping subarrays of dimension $M-L+1$. At each time $n$, $L$ snapshots of dimension $M-L+1$ are thus available, and the scheme
provides $NL$ observations of dimension $M-L+1$. In order
to be more specific, we introduce the following notations. If $L$ is an integer less than $M$,
we denote by $\mathcal{Y}_n^{(L)}$ the $(M - L+1) \times L$ Hankel matrix defined by
\begin{equation}
\label{eq:def-calXL}
\mathcal{Y}_n^{(L)} = \left( \begin{array}{ccccc} {\bf y}_{1,n} & {\bf y}_{2,n} & \ldots & \ldots & {\bf y}_{L,n} \\
                                                  {\bf y}_{2,n}  & {\bf y}_{3,n}  & \ldots & \ldots & {\bf y}_{L+1,n} \\
                                                   \vdots &  \vdots &   \vdots &   \vdots &   \vdots   \\
                                                   \vdots &  \vdots &   \vdots &   \vdots &   \vdots   \\
                                                   {\bf y}_{M-L+1,n} & {\bf y}_{M-L+2,n} & \ldots & \ldots & {\bf y}_{M,n} \end{array} \right)
\end{equation}
Column $l$ of matrix $\mathcal{Y}_n^{(L)}$ corresponds to the observation on subarray $l$ at time $n$.
Collecting all the observations on the various subarrays allows to obtain $NL$ snapshots, thus increasing
artificially the number of observations. We define ${\bf Y}_N^{(L)}$ as the $(M-L+1) \times NL$ block-Hankel matrix given by
\begin{equation}
\label{eq:def-YNL}
{\bf Y}_N^{(L)} = \left( \mathcal{Y}_1^{(L)}, \ldots, \mathcal{Y}_N^{(L)} \right)
\end{equation}
In order to express ${\bf Y}_N^{(L)}$, we consider the $(M-L+1) \times L$ Hankel matrix
$\mathcal{A}^{(L)}(\theta)$ defined from vector ${\bf a}_M(\theta)$ in the same way than
$\mathcal{Y}_n^{(L)}$. We remark that $\mathcal{A}^{(L)}(\theta)$
is rank 1, and can be written as
\begin{equation}
\label{eq:expre-calAtheta}
\mathcal{A}^{(L)}(\theta) = \sqrt{L(M-L+1)/M} \; {\bf a}_{M-L+1}(\theta) \, ({\bf a}_{L}(\theta))^{T}
\end{equation}
We consider the $(M-L+1) \times KL$ matrix ${\bf A}^{(L)}$
\begin{equation}
\label{eq:def-YNL}
{\bf A}^{(L)} = \left( \mathcal{A}^{(L)}(\theta_1), \mathcal{A}^{(L)}(\theta_2), \ldots, \mathcal{A}^{(L)}(\theta_K) \right)
\end{equation}
which, of course, is a rank $K$ matrix whose range coincides with the
subspace generated by the $(M-L+1)$-dimensional vectors ${\bf a}_{M-L+1}(\theta_1), \ldots, {\bf a}_{M-L+1}(\theta_K)$. ${\bf Y}_N^{(L)}$ can be written as
\begin{equation}
\label{eq:expre-YNL}
{\bf Y}_N^{(L)} = {\bf A}^{(L)} \, \left( {\bf S}_N \otimes {\bf I}_L \right) + {\bf V}_N^{(L)}
\end{equation}
where matrix ${\bf V}_N^{(L)}$ is the block Hankel matrix corresponding to the additive noise. As matrix ${\bf S}_N \otimes {\bf I}_L$ is full rank, the extended obervation matrix ${\bf Y}_N^{(L)}$
appears as a noisy version of a low rank component whose range is the
$K$--dimensional subspace generated by vectors  ${\bf a}_{M-L+1}(\theta_1), \ldots, {\bf a}_{M-L+1}(\theta_K)$. Moreover, it is easy to check that
$$
\mathbb{E} \left( \frac{{\bf V}_N^{(L)}{\bf V}_N^{(L)*}}{NL} \right) = \sigma^{2} {\bf I}_{M-L+1}
$$
Therefore, it is potentially possible to estimate the DoAs $(\theta_k)_{k=1, \ldots, K}$
using a subspace approach based on the eigenvalues / eigenvectors decomposition of matrix ${\bf Y}_N^{(L)} {\bf Y}_N^{(L)*} / NL$. The asymptotic behaviour of spatial smoothing subspace methods is standard in the regimes where $M-L+1$ remains fixed while $NL$ converges towards $\infty$. This is due to the law of large numbers which implies that
the empirical covariance matrix ${\bf Y}_N^{(L)}{\bf Y}_N^{(L)*}/ NL$  has the
same asymptotic behaviour than  ${\bf A}^{(L)} \,  \left( {\bf S}_N {\bf S}_N^{*} \otimes {\bf I}_L / NL \right) \; {\bf A}^{(L)*}  + \sigma^{2} {\bf I}_{M-L+1}$,
In this context, the orthogonal projection matrix $\hat{{\bs \Pi}}_N^{(L)}$ onto the eigenspace associated to the $M-L+1-K$ smallest eigenvalues of ${\bf Y}_N^{(L)}{\bf Y}_N^{(L)*}/ NL$ is a consistent estimate of the orthogonal projection matrix ${\bs \Pi}^{(L)}$ on the noise subspace, i.e. the orthogonal complement of $\mathrm{sp} \{ {\bf a}_{M-L+1}(\theta_1), \ldots, {\bf a}_{M-L+1}(\theta_K) \}$. In other words, it holds that
\begin{equation}
\label{eq:consistance-Pi}
\left\| \hat{{\bs \Pi}}_N^{(L)} - {\bs \Pi}^{(L)} \right\| \rightarrow 0 \; a.s.
\end{equation}
The traditional pseudo-spectrum estimate $\hat{\eta}_N^{(t)}(\theta)$ defined by
$$
\hat{\eta}_N^{(t)}(\theta) = {\bf a}_{M-L+1}(\theta)^{*} \hat{{\bs \Pi}}_N^{(L)} {\bf a}_{M-L+1}(\theta)
$$
thus verifies
\begin{align}
	\sup_{\theta \in [-\pi,\pi]} \left|\hat{\eta}_N^{(t)}(\theta)-\eta(\theta)\right| \xrightarrow[N \to \infty]{a.s.} 0.
\label{eq:consistency-trad-pseudo-spectrum}
\end{align}
where $\eta(\theta) = {\bf a}_{M-L+1}(\theta)^{*} {\bs \Pi}^{(L)} {\bf a}_{M-L+1}(\theta)$ is the MUSIC pseudo-spectrum. Moreover,
the $K$ MUSIC traditional DoA estimates, defined formally, for $k=1,\ldots,K$, by
\begin{equation}
\label{eq:def-theta-trad}
	\hat{\theta}^{(t)}_{k,N} = \argmin_{\theta \in \Ical_k} \hat{\eta}_N^{(t)}(\theta),
\end{equation}
where $\Ical_k$ is a compact interval containing $\theta_k$ and such that $\Ical_k \cap \Ical_l = \emptyset$ for $k \neq l$,
are consistent, i.e.
\begin{align}
	\hat{\theta}^{(t)}_{k,N} \xrightarrow[N \to \infty]{a.s.} \theta_k.
	\label{eq:consistency-DoA-traditional}
\end{align}

However, the regime where $M-L+1$ remains fixed while $NL$ converges towards $\infty$ is not very interesting in practice because the size $M-L+1$ of the subarrays may be much smaller that the number of antennas $M$, thus reducing the resolution of the method. We therefore study spatial smoothing schemes in regimes where the dimensions $M-L+1$ and $NL$ of matrix ${\bf Y}_N^{(L)}$ are of the same order
of magnitude and where $\frac{L}{M} \rightarrow 0$ in order to keep unchanged the
aperture of the array. More precisely, we assume that integers $N$ and $L$ depend
on $M$ and that
\begin{equation}
\label{eq:regime-M-N-L}
\hspace{-0.5cm}M \rightarrow +\infty, N = \mathcal{O}(M^{\beta}), \;  \frac{1}{3} < \beta  \leq 1 , \;  c_N = \frac{M-L+1}{NL} \rightarrow c_*
\end{equation}


In regime (\ref{eq:regime-M-N-L}), $N$ thus converges towards $\infty$ but at a rate that may be much lower than $M$ thus modelling contexts in which $N$ is much smaller than $M$. As $N \rightarrow +\infty$, it also holds that $\frac{M}{NL} \rightarrow c_*$. Therefore, it is clear that $L = \mathcal{O}(M^{\alpha})$ where $\alpha = 1 - \beta$ verifies
with $0 \leq \alpha < 2/3$. $L$ may thus converge towards $\infty$ (even faster than $N$ if $\beta < 1/2$) but in such a way that $\frac{L}{M}\rightarrow 0$. As in regime (\ref{eq:regime-M-N-L}) $N$ depends on $M$,
it could be appropriate to index the various matrices and DoA estimators by integer $M$ rather than by integer $N$ as in definitions (\ref{eq:def-YNL}) and (\ref{eq:def-theta-trad}). However, we prefer to use the index $N$ in the following
in order to keep the notations unchanged. We also denote projection matrix $\bs{\Pi}^{(L)}$ and pseudo-spectrum $\eta(\theta)$ by  $\bs{\Pi}_N^{(L)}$ and $\eta_N(\theta)$ because they depend on
$M$. Moreover, in the following, the notation $N \rightarrow +\infty$
should be understood as regime (\ref{eq:regime-M-N-L}) for some $\beta \in (1/3, 1]$.

\subsection{Main results.}
In regime (\ref{eq:regime-M-N-L}), (\ref{eq:consistance-Pi}) is no more valid. Hence, (\ref{eq:consistency-DoA-traditional}) is questionable. In this paper, we show that it is possible to generalize the G-MUSIC  estimators
introduced in \cite{hachem2012subspace} in the case where $L=1$ to the context of
spatial smoothing schemes in regime (\ref{eq:regime-M-N-L}) and establish the following results. Under the separation condition that the $K$ non zero eigenvalues of matrix $\frac{1}{L} {\bf A}^{(L)} \left( \frac{{\bf S}_N{\bf S}_N^{*}}{N} \otimes {\bf I}_L \right) {\bf A}^{(L)*}$ are above the threshold $\sigma^{2} \sqrt{c_*}$ for each $N$ large enough, we deduce from \cite{vallet-loubaton-mestre-2014} that:
\begin{itemize}
\item the spatial smoothing traditional MUSIC estimates  $(\hat{\theta}_{k,N}^{(t)})_{k=1, \ldots, K}$  and the G-MUSIC SS estimates, denoted
$(\hat{\theta}_{k,N})_{k=1, \ldots, K}$ are consistent and verify
\begin{eqnarray}
\label{eq:M-consistency-t}
M (\hat{\theta}_{k,N}^{(t)} - \theta_k) & \rightarrow & 0 \; a.s., \\
\label{eq:M-consistency-G}
 M (\hat{\theta}_{k,N} - \theta_k) & \rightarrow & 0 \; a.s.
\end{eqnarray}
\end{itemize}
(\ref{eq:M-consistency-t}) and (\ref{eq:M-consistency-G}) hold when
the DoA $(\theta_k)_{k=1, \ldots, K}$ do not scale with $M,N$. In pratice,
this assumption corresponds to practical situations where the DoA are widely
spaced because when the DoA $(\theta_k)_{k=1, \ldots, K}$ are fixed, the ratio
$$
\frac{\min_{k \neq l} |\theta_k - \theta_l|}{(2 \pi)/M}
$$
converges towards $\infty$. We deduce from \cite{vallet-loubaton-mestre-2014} that:
\begin{itemize}
\item If $K=2$ and that the 2 DoAs scale with $M,N$ is such a way that $\theta_{2,N} - \theta_{1,N} = \mathcal{O}(\frac{1}{M})$, then the G-MUSIC SS estimates
still verify (\ref{eq:M-consistency-G}) while the traditional MUSIC SS estimates no longer verify
(\ref{eq:M-consistency-t})
\end{itemize}
As in the case $L=1$, the above mentioned separation condition ensures that
the $K$ largest eigenvalues of the empirical covariance matrix $({\bf Y}_N^{(L)} {\bf Y}_N^{(L)*})/NL$
correspond to the $K$ sources, and the signal and noise subspaces can be separated. In order to
obtain some insights on this condition, and on the potential benefit of the spatial smoothing,
we study the separation condition when
$M$ and $N$ converge towards $\infty$ at the same rate, i.e. when $\frac{M}{N} \rightarrow d_*$,
or equivalently that $\beta = 1$ and that $L$ does not scale with $N$.
In this case, it is clear  that $c_*$ coincides
with $c_* = d_*/L$. Under the assumption that $\frac{{\bf S}_N{\bf S}_N^{*}}{N}$ converges towards a diagonal matrix ${\bf D}$ when $N$ increases, then we establish that the separation condition holds if
\begin{equation}
\label{eq:separation-condition-spatial-smoothing}
\lambda_{K} \left( {\bf A}_{M-L+1}^{*}  {\bf A}_{M-L+1} {\bf D}  \right) > \frac{\sigma^{2} \sqrt{d_*}}{\sqrt{L}}
\end{equation}
for each $(M,N)$ large enough. If $L=1$, the separation condition introduced in the context of (unsmoothed) G-MUSIC algorithms
(\cite{hachem2012subspace}) is of course recovered, i.e.
$$
\lambda_{K} \left( {\bf A}_{M}^{*}  {\bf A}_{M} {\bf D}  \right) > \sigma^{2} \sqrt{d_*}
$$
If $M$ is large and that $L << M$, matrix ${\bf A}_{M-L+1}^{*}  {\bf A}_{M-L+1}$ is close from
 ${\bf A}_{M}^{*}  {\bf A}_{M}$ and the separation condition is nearly equivalent to
$$
\lambda_{K} \left( {\bf A}_{M}^{*}  {\bf A}_{M} {\bf D}  \right) > \frac{\sigma^{2} \sqrt{d_*}}{\sqrt{L}}
$$
Therefore, it is seen that the use of the spatial smoothing scheme allows to reduce the threshold $\sigma^{2} \sqrt{d_*}$ corresponding to G-MUSIC method without spatial smoothing by the factor $\sqrt{L}$.  Therefore, if
$M$ and $N$ are the same order of magnitude, our asymptotic analysis allows to predict an improvement of the performance of the G-MUSIC SS methods when $L$ increases provided $L << M$. If $L$  becomes too large, the above rough analysis is no more justified and the impact of the diminution of the number of antennas becomes dominant, and the performance tends to decrease.

\section{Asymptotic behaviour of the largest singular values and corresponding singular vectors
of finite rank perturbations of certain large random block-Hankel matrices.}
\label{sec:technical-results}
In this section, $N,M,L$ still satisfy (\ref{eq:regime-M-N-L}) while $K$ is a fixed integer
that does not scale with $N$. We consider the $(M+L-1) \times NL$ block-Hankel random
matrix ${\bf V}_N^{(L)}$ defined previously,
and introduce matrix ${\bf Z}_N$ defined
$${\bf Z}_N = \frac{1}{\sqrt{NL}} {\bf V}_N^{(L)}$$
in order to simplify the notations. The entries of ${\bf Z}_N$ have of course variance
$\sigma^{2}/NL$. In the following, ${\bf B}_N$ represents a deterministic $(M+L-1) \times NL$
matrix verifying
\begin{equation}
\label{eq:B-bounded-B-rank}
\hspace{-0.5cm}\sup_N \|{\bf B}_N \| < +\infty, \; \mathrm{Rank}({\bf B}_N) = K,
\end{equation}
for each $N$ large enough.
We denote by $\lambda_{1,N} > \lambda_{2,N} \ldots > \lambda_{K,N}$ the non zero eigenvalues of matrix ${\bf B}_N {\bf B}_N^{*}$ arranged in decreasing order, and by $({\bf u}_{k,N})_{k=1, \ldots, K}$
and $(\tilde{{\bf u}}_{k,N})_{k=1, \ldots, K}$ the associated left and right singular vectors
of ${\bf B}_N$. The singular value decomposition of ${\bf B}_N$ is thus given by
$$
{\bf B}_N = \sum_{k=1}^{K} \lambda_{k,N}^{1/2} {\bf u}_{k,N} \tilde{{\bf u}}_{k,N}^{*} =
{\bf U}_N {\bs \Lambda}_N^{1/2} \tilde{{\bf U}}_N^{*}
$$
Moreover, we assume that:
\begin{assum}
\label{as:convergence-valeurs-propres-B}
The $K$ non zero eigenvalues $(\lambda_{k,N})_{k=1, \ldots, K}$ of matrix ${\bf B}_N {\bf B}_N^{*}$ converge towards $\lambda_1 > \lambda_2 > \ldots > \lambda_K$ when $N \rightarrow +\infty$.
\end{assum}
Here, for ease of exposition, we assume that the eigenvalues $(\lambda_{k,N})_{k=1, \ldots, K}$ have multiplicity 1 and that $\lambda_k \neq \lambda_l$ for $k \neq l$. However, the forthcoming
results can be easily adapted if some $\lambda_k$ coincide.

We define matrix ${\bf X}_N$ as
\begin{equation}
\label{eq:def-X}
{\bf X}_N = {\bf B}_N + {\bf Z}_N
\end{equation}
${\bf X}_N$ can thus be interpreted as a rank $K$ perturbation of the block-Hankel
matrix ${\bf Z}_N$. The purpose of this section is to study the behaviour of
the $K$ largest eigenvalues $(\hat{{\lambda}}_{k,N})_{k=1, \ldots, K}$ of matrix
${\bf X}_N {\bf X}_N^{*}$ as well as of their corresponding eigenvectors
$(\hat{{\bf u}}_{k,N})_{k=1, \ldots, K}$.
It turns out that $(\hat{{\lambda}}_{k,N})_{k=1, \ldots, K}$
and $(\hat{{\bf u}}_{k,N})_{k=1, \ldots, K}$ behave as if the entries of matrix ${\bf Z}_N$ where
i.i.d. To see this, we have first to precise the behaviour of the eigenvalues of matrix ${\bf Z}_N {\bf Z}_N^{*}$ in the asymptotic regime (\ref{eq:regime-M-N-L}).

\subsection{Behaviour of the eigenvalues of matrix ${\bf Z}_N {\bf Z}_N^{*}$.}
We first recall the definition of the Marcenko-Pastur distribution $\mu_{\sigma^{2}, c}$ of parameters
$\sigma^{2}$ and $c$ (see e.g. \cite{bai-silverstein-book}). $\mu_{\sigma^{2}, c}$ is the probability distribution
defined by
\begin{align}
d\mu_{\sigma^{2},c}(x) = \delta_0 [1-c^{-1}]_{+} \, + \frac{\sqrt{\left(x - x^-\right)\left(x^+ - x\right)}}{2 \sigma^2 c \pi x} \mathbb{1}_{[x^-,x^+]}(x) \, dx
	\notag
\end{align}
with $x^- = \sigma^2 (1-\sqrt{c})^2$ and $x^+ = \sigma^2 (1+\sqrt{c})^2$. Its Stieltjes transform
$m_{\sigma^{2},c}(z)$ defined by
\begin{align}
	 m_{\sigma^{2},c}(z) = \int_{\Rbb} \frac{\drm \mu_{\sigma^{2},c} (\lambda)}{\lambda - z}
	\notag
\end{align}
is known to satisfy the fundamental equation
\begin{align}
\label{eq:fundamental_equation_t}
	m_{\sigma^{2},c}(z) = \frac{1}{-z + \sigma^2 \frac{1}{1 + \sigma^{2} c m_{\sigma^{2},c}(z)}}
\end{align}
or equivalently,
\begin{align}
\label{eq:fundamental_equation_t-ttilde-1}
	m_{\sigma^{2},c}(z) = \frac{1}{-z(1 + \sigma^2 \tilde{m}_{\sigma^{2},c}(z))} \\
\label{eq:fundamental_equation_t-ttilde-2}
        \tilde{m}_{\sigma^{2},c}(z) = \frac{1}{-z(1 + \sigma^2 c m_{\sigma^{2},c}(z))}
\end{align}
where $\tilde{m}_{\sigma^{2},c}(z)$ is known to coincide with Stieltjes transform of the
Marcenko-Pastur distribution $\mu_{\sigma^{2}c, c^{-1}} = c \mu_{\sigma^{2}, c} + (1-c) \delta_0$.

In order to simplify the notations, we denote by $m_*(z)$ and $\tilde{m}_*(z)$ the Stieltjes transforms of Marcenko-Pastur distributions $\mu_{\sigma^{2},c_*}$ and $\mu_{\sigma^{2}c_*, c_*^{-1}}$.
$m_*(z)$ and $\tilde{m}_*(z)$ verify Equations (\ref{eq:fundamental_equation_t-ttilde-1})
and (\ref{eq:fundamental_equation_t-ttilde-2}) for $c=c_*$. We also denote by $x_*^{-}$ and $x_*^{+}$
the terms $x_{*}^{-} =  \sigma^{2}(1 - \sqrt{c_*})^{2}$ and $x_{*}^{+} =  \sigma^{2}(1 + \sqrt{c_*})^{2}$.
We recall that function $w_*(z)$ defined by
\begin{equation}
\label{eq:def-w}
w_{*}(z)  = \frac{1}{z \, m_*(z) \, \tilde{m}_*(z)}
\end{equation}
is analytic on $\mathbb{C} - [x_{*}^{-}, x_{*}^{+}]$, verifies $w_*(x_{*}^{+}) = \sigma^{2} \sqrt{c_*}$,
and increases from $\sigma^{2} \sqrt{c_*}$ to $+\infty$ when $x$ increases from $x_{*}^{+}$ to
$+\infty$ (see \cite{benaych2012singular}, section 3.1). Moreover, if $\phi_*(w)$ denotes function defined by
\begin{equation}
\label{eq:def-phi}
\phi_*(w) = \frac{(w+\sigma^2)(w+\sigma^2c_*)}{w}
\end{equation}
then, $\phi_*$ increases from  $x_{*}^{+}$ to $+\infty$ when $w$ increases from $\sigma^{2} \sqrt{c_*}$ to $+\infty$. Finally, it holds that
\begin{equation}
\label{eq:phi-inverse-w}
\phi_*(w_*(z)) = z
\end{equation}
for each $z \in \mathbb{C} - [x_{*}^{-}, x_{*}^{+}]$.

We denote by
${\bf Q}_N(z)$ and $\tilde{\Q}_N(z)$ the so-called resolvent of matrices
${\bf Z}_N {\bf Z}_N^{*}$ and ${\bf Z}_N^{*} {\bf Z}_N$ defined by
$$
{\bf Q}_N(z) = \left( {\bf Z}_N {\bf Z}_N^{*} - z {\bf I}_{M-L+1} \right)^{-1}, \;
\tilde{{\bf Q}}_N(z) = \left( {\bf Z}_N^* {\bf Z}_N - z {\bf I}_{NL} \right)^{-1}
$$
Then, the results of \cite{loubaton-bloc-hankel} imply the following proposition.
\begin{proposition}
\label{prop:eigenvalues-ZZ*}
\begin{itemize}
\item
(i) The eigenvalue distribution of matrix ${\bf Z}_N {\bf Z}_N^{*}$ converges almost
surely towards the Marcenko-Pastur distribution $\mu_{\sigma^{2}, c_*}$, or equivalently,
for each $z \in \mathbb{C} - \mathbb{R}^{+}$,
\begin{equation}
\label{eq:convergence-MP-ZZ*}
\frac{1}{M-L+1} \mathrm{Tr}({\bf Q}_N(z)) - m_*(z) \rightarrow 0 \, a.s.
\end{equation}
\item
(ii) For each $\epsilon > 0$, almost surely, for $N$ large enough, all the eigenvalues of ${\bf Z}_N {\bf Z}_N^{*}$ belong to $[\sigma^{2}(1 - \sqrt{c_*})^{2} - \epsilon, \sigma^{2}(1 + \sqrt{c_*})^{2} + \epsilon]$
if $c_* \leq 1$, and to $[\sigma^{2}(1 - \sqrt{c_*})^{2} - \epsilon, \sigma^{2}(1 + \sqrt{c_*})^{2} + \epsilon] \cup \{0 \}$ if $c_* > 1$.
\item (iii)
Moreover, if ${\bf a}_N, {\bf b}_N$ are  $(M-L+1)$--dimensional deterministic
vectors satisfying $\sup_{N} (\|{\bf a}_N \|, \|{\bf b}_N \|) < +\infty$ , then it holds that for each $z \in \mathbb{C}^{+}$
\begin{equation}
\label{eq:convergence-forme-quadratique-Q}
{\bf a}_N^{*} \left( {\bf Q}_N(z) - m_*(z) {\bf I} \right) {\bf b}_N \rightarrow 0 \; a.s.
\end{equation}
Similarly, if $\tilde{{\bf a}}_N$ and $\tilde{{\bf b}}_N$ are  $NL$--dimensional deterministic
vectors verifying $\sup_{N} (\|\tilde{{\bf a}}_N\|, \|\tilde{{\bf b}}_N\|) < +\infty$, then for each $z \in \mathbb{C}^{+}$, it holds that
\begin{equation}
\label{eq:convergence-forme-quadratique-tildeQ}
\tilde{{\bf a}}_N^{*} \left( \tilde{{\bf Q}}_N(z) - \tilde{m}_*(z) {\bf I} \right) \tilde{{\bf b}}_N \rightarrow 0 \; a.s.
\end{equation}
Moreover, for each $z \in \mathbb{C}^{+}$, it holds that
\begin{equation}
\label{eq:convergence-terme-mixte}
{\bf a}_N^{*} \left( {\bf Q}_N(z) {\bf Z}_N \right) {\bf b}_N \rightarrow 0 \; a.s.
\end{equation}
Finally, for each $\epsilon > 0$, convergence properties (\ref{eq:convergence-forme-quadratique-Q}, \ref{eq:convergence-forme-quadratique-tildeQ}, \ref{eq:convergence-terme-mixte}) hold uniformly w.r.t. $z$ on each compact subset of $\mathbb{C} - [0, x_*^{+} + \epsilon]$.
\end{itemize}
\end{proposition}
The proof is given in the Appendix.

\begin{remark}
Proposition \ref{prop:eigenvalues-ZZ*} implies that
in a certain sense, matrix ${\bf Z}_N {\bf Z}_N^{*}$ behaves as if the entries of
${\bf Z}_N$ were i.i.d because Proposition \ref{prop:eigenvalues-ZZ*} is known to hold for
i.i.d. matrices. In the i.i.d. case, (\ref{eq:convergence-MP-ZZ*}) was established for
the first time in \cite{marcenko1967distribution}, the almost sure location
of the eigenvalues of ${\bf Z}_N {\bf Z}_N^{*}$ can be found in \cite{bai-silverstein-book} (see
Theorem 5-11), while (\ref{eq:convergence-forme-quadratique-Q}), (\ref{eq:convergence-forme-quadratique-tildeQ}) and (\ref{eq:convergence-terme-mixte}) are trivial modifications of
Lemma 5 of \cite{hachem2012subspace}.
\end{remark}

We notice that the convergence towards the Marcenko-Pastur
distribution holds as soon as $N \rightarrow +\infty$ and $\frac{M-L+1}{NL} \rightarrow c_*$.
In particular, the convergence is still valid if $N = \mathcal{O}(M^{\beta})$
for each $0 < \beta \leq 1$, or equivalently if $L = \mathcal{O}(M^{\alpha})$ for each $0 \leq \alpha < 1$.
$L$ can therefore converge towards $\infty$ much faster than $N$. However, the hypothesis that $\beta > 1/3$, which is also equivalent to $L = \mathcal{O}(M^{\alpha})$ with $\alpha < 2/3$, is necessary to establish
item (ii).
\subsection{The $K$ largest eigenvalues and eigenvectors of
${\bf X}_N {\bf X}_N^{*}$.}
While matrix ${\bf Z}_N$ does not meet the conditions formulated
in \cite{benaych2012singular}, Proposition \ref{prop:eigenvalues-ZZ*} allows to
use the approach used in \cite{benaych2012singular}, and to prove that
the $K$ largest eigenvalues and corresponding eigenvectors of ${\bf X}_N {\bf X}_N^{*}$.
behave as if the entries of ${\bf Z}_N$ were i.i.d. In particular, the following result holds.
\begin{theorem}
	\label{theorem:spiked_eig}
	We denote by $s$, $0 \leq s \leq K$, the largest integer for which
\begin{equation}
\label{eq:seuil}
\lambda_s > \sigma^{2} \sqrt{c_*}
\end{equation}
Then, for $k=1,\ldots,s$, it holds that
	\begin{align}
        \label{eq:convergence-lambda-spike}
		\hat{\lambda}_{k,N} \xrightarrow[N\to\infty]{a.s.} \rho_k = \phi(\lambda_k) = \frac{(\lambda_k + \sigma^2)(\lambda_k + \sigma^2 c)}{\lambda_k} > x_*^{+}.
	\end{align}
	 Moreover, for $k=s+1, \ldots, K$,
it holds that
	\begin{align}
        \label{eq:converge-lambda-bulk}
		\hat{\lambda}_{k,N} \rightarrow \sigma^{2}(1+\sqrt{c_*})^{2} \, a.s.
	\end{align}
Finally, for all deterministic sequences of unit norm vectors $(\d_{1,N})$, $(\d_{2,N})$, we have for $k=1,\ldots,s$
	\begin{align}
\label{eq:convergence-vecteurs-propres}
		&\d_{1,N}^* \hat{\u}_{k,N} \hat{\u}_{k,N}^* \d_{2,N} =
		\notag \\
		&\qquad\qquad h_*(\rho_k) \d_{1,N}^* \u_{k,N} \u_{k,N}^* \d_{2,N} + o(1) \quad a.s.,
	\end{align}
where function $h_*(z)$ is defined by
\begin{equation}
\label{eq:def-h*}
h_*(z) =  \frac{w_*(z)^2 - \sigma^4 c_*}{w_*(z) (w_*(z) + \sigma^2 c_*)}
\end{equation}
\end{theorem}
For the reader's convenience, we provide in the appendix some insights on the approach developed in
\cite{benaych2012singular} to prove (\ref{eq:convergence-lambda-spike}) and (\ref{eq:converge-lambda-bulk}). For more details on (\ref{eq:convergence-vecteurs-propres}), see the proof of Theorem 2 in \cite{hachem2012subspace}
as well as the identity
$$
h_*(z) = \frac{z m_*(z)^{2} \tilde{m}_*(z)}{(z m_*(z) \tilde{m}_*(z))^{'}}
$$
where $'$ represents the derivation w.r.t. $z$.

\section{Derivation of a consistent G-MUSIC method.}
\label{sec:utilisation}
We now use the results of section \ref{sec:technical-results} for
matrix ${\bf X}_N = {\bf Y}_N^{(L)} / \sqrt{NL}$ and ${\bf B}_N = \frac{1}{\sqrt{NL}} {\bf A}^{(L)} ({\bf S}_N \otimes {\bf I}_L)$. We recall that
$(\hat{\lambda}_{k,N})_{k=1, \ldots, M-L+1}$ and $(\hat{{\bf u}}_{k,N})_{k=1, \ldots, M-L+1}$ represent the
eigenvalues and eigenvectors of the empirical covariance matrix ${\bf Y}_N^{(L)} {\bf Y}_N^{(L)*}/ NL$,
and that $(\lambda_{k,N})_{k=1, \ldots, K}$ and $({\bf u}_{k,N})_{k=1, \ldots, K}$ are the non zero eigenvalues and corresponding eigenvectors of $\frac{1}{L} {\bf A}^{(L)} \left({\bf S}_N{\bf S}_N^{*}/N \otimes {\bf I}_L \right)  {\bf A}^{(L)*}$.
We recall that ${\bs \Pi}_N^{(L)}$ represents the orthogonal projection matrix onto the noise subspace, i.e. the orthogonal complement of the space generated by vectors $({\bf a}_{M-L+1}(\theta_k))_{k=1, \ldots, K}$
and that $\eta_N(\theta)$ is the corresponding MUSIC pseudo-spectrum
$$
\eta_N(\theta) = {\bf a}_{M-L+1}(\theta)^{*} \, {\bs \Pi}_N^{(L)} \, {\bf a}_{M-L+1}(\theta)
$$
Theorem \ref{theorem:spiked_eig} allows to generalize immediately the results of
\cite{hachem2012subspace} and \cite{vallet-loubaton-mestre-2014} concerning the
consistency of G-MUSIC and MUSIC DoA estimators in the case $L=1$. More precisely:
\begin{theorem}
\label{thm:separation-spatial-smoothing}
Assume that the $K$ non zero eigenvalues $(\lambda_{k,N})_{k=1, \ldots, K}$ converge towards deterministic terms $\lambda_1 > \lambda_2 > \ldots >\lambda_K$ and that
\begin{equation}
\label{eq:separation}
\lambda_K > \sigma^{2} \sqrt{c_*}
\end{equation}
Then, the estimator $\hat{\eta}_N(\theta)$ of the pseudo-spectrum $\eta_N(\theta)$ defined by
\begin{align}
	\hat{\eta}_N(\theta)  = ({\bf a}_{M-L+1}(\theta))^* \left(\I - \sum_{k=1}^K \frac{1}{h\left(\hat{\lambda}_{k,N} \right)} \hat{\u}_{k,N} \hat{\u}_{k,N}^*\right) {\bf a}_{M-L+1}(\theta)
	\label{eq:noise_subspace_estimator}
\end{align}
verifies
\begin{align}
	\sup_{\theta \in [-\pi,\pi]} \left|\hat{\eta}_N(\theta)-\eta_N(\theta)\right| \xrightarrow[N \to \infty]{a.s.} 0,
	\label{eq:uniform_consistency}
\end{align}
\end{theorem}
This result can be proved as Proposition 1 in \cite{hachem2012subspace}.

In order to
obtain some insights on condition (\ref{eq:separation}) and on the potential benefits of the spatial smoothing,
we explicit the separation condition (\ref{eq:separation}) when
$M$ and $N$ converge towards $\infty$ at the same rate, i.e. when $\frac{M}{N} \rightarrow d_*$,
or equivalently that $\beta = 1$ and that $L$ does not scale with $N$.
In this case, it is clear  that $c_*$ coincides
with $c_* = d_*/L$. It is easily seen that
\begin{equation}
\label{eq:expre-signal-empirical-covariance-matrix}
\frac{1}{L} {\bf A}^{(L)} \left( \frac{{\bf S}_N{\bf S}_N^{*}}{N} \otimes {\bf I}_L \right) {\bf A}^{(L)*} =
\left(M-L+1/M \right) \; {\bf A}_{M-L+1} \left( \frac{{\bf S}_N{\bf S}_N^{*}}{N} \bullet {\bf A}_L^{T} \overline{{\bf A}}_L \right) {\bf A}_{M-L+1}^{*}
\end{equation}
where $\bullet$ represents the Hadamard (i.e. element wise) product of matrices, and where $\overline{{\bf B}}$ stands for the complex conjugation operator of the elements of matrix ${\bf B}$.
If we assume that $\frac{{\bf S}_N{\bf S}_N^{*}}{N}$ converges towards a diagonal matrix ${\bf D}$ when $N$ increases, then $\frac{{\bf S}_N{\bf S}_N^{*}}{N} \bullet ({\bf A}_L^{T} \overline{{\bf A}}_L)$
converges towards the diagonal matrix ${\bf D} \bullet \mathrm{Diag} \left( {\bf A}_L^{T} \overline{{\bf A}}_L \right) = {\bf D}$. Therefore, $\frac{{\bf S}_N{\bf S}_N^{*}}{N} \bullet ({\bf A}_L^{T} \overline{{\bf A}}_L) \simeq  {\bf D}$ when is large enough. Using that $\frac{L}{M} \rightarrow 0$, we obtain that  the separation condition is nearly equivalent to
$$
\lambda_{K} \left( {\bf A}_{M-L+1} {\bf D} \;  {\bf A}_{M-L+1}^{*} \right) > \frac{\sigma^{2} \sqrt{d_*}}{\sqrt{L}}
$$
or to
\begin{equation}
\label{eq:separation-condition-spatial-smoothing}
\lambda_{K} \left( {\bf A}_{M-L+1}^{*}  {\bf A}_{M-L+1} {\bf D}  \right) > \frac{\sigma^{2} \sqrt{d_*}}{\sqrt{L}}
\end{equation}
for each $(M,N)$ large enough. If $L=1$, the separation condition introduced in the context of (unsmoothed) G-MUSIC algorithms
(\cite{hachem2012subspace}) is of course recovered, i.e.
$$
\lambda_{K} \left( {\bf A}_{M}^{*}  {\bf A}_{M} {\bf D}  \right) > \sigma^{2} \sqrt{d_*}
$$
for each $(M,N)$ large enough. If $M$ is large and that $L << M$, matrix ${\bf A}_{M-L+1}^{*}  {\bf A}_{M-L+1}$ is close from
 ${\bf A}_{M}^{*}  {\bf A}_{M}$ and the separation condition is nearly equivalent to
$$
\lambda_{K} \left( {\bf A}_{M}^{*}  {\bf A}_{M} {\bf D}  \right) > \frac{\sigma^{2} \sqrt{d_*}}{\sqrt{L}}
$$
Therefore, it is seen that the use of the spatial smoothing scheme allows to reduce the threshold $\sigma^{2} \sqrt{d_*}$ corresponding to G-MUSIC method without spatial smoothing by the factor $\sqrt{L}$.  Hence, if
$M$ and $N$ are the same order of magnitude, our asymptotic analysis allows to predict an improvement of the performance of the G-MUSIC methods based on spatial smoothing when $L$ increases provided $L << M$. If $L$  becomes too large, the above rough analysis is no more justified and the impact of the diminution of the number of antennas becomes dominant, and the performance tends to decrease. This analysis is sustained by the numerical simulations presented in section \ref{sec:simulation}. \\

We define the
DoA G-MUSIC SS estimates $(\hat{\theta}_{k,N})_{k=1, \ldots, K}$ by
\begin{equation}
\label{eq:def-theta-GMUSIC}
	\hat{\theta}_{k,N} = \argmin_{\theta \in \Ical_k}  \left| \hat{\eta}_N(\theta) \right|,
\end{equation}
where $\Ical_k$ is a compact interval containing $\theta_k$ and such that $\Ical_k \cap \Ical_l = \emptyset$ for $k \neq l$. As in \cite{hachem2012subspace}, (\ref{eq:uniform_consistency})
as well as the particular structure of directional vectors ${\bf a}_{M-L+1}(\theta)$
imply the following result which can be proved as Theorem 3 of \cite{hachem2012subspace}
\begin{theorem}
\label{th:consistency-GMUSIC-DoA-estimates}
Under condition (\ref{eq:separation}), the DoA G-MUSIC SS estimates $(\hat{\theta}_{k,N})_{k=1, \ldots, K}$
verify
\begin{equation}
\label{eq:M-consistency-GMUSIC-DoA-estimates}
M \left( \hat{\theta}_{k,N} - \theta_k \right) \rightarrow 0 \; a.s.
\end{equation}
for each $k=1, \ldots, K$.
\end{theorem}
\begin{remark}
We remark that under the extra assumption that $\frac{{\bf S}_N{\bf S}_N^{*}}{N}$ converges towards a diagonal matrix,\cite{hachem2012subspace} (see also \cite{vallet-mestre-loubaton-eusipco-2012} for more general matrices ${\bf S}$) proved when $L = 1$ that $M^{3/2} \left( \hat{\theta}_{k,N} - \theta_k \right)$ converges in distribution towards a Gaussian distribution. It would be interesting
to generalize the results of \cite{hachem2012subspace} and \cite{vallet-mestre-loubaton-eusipco-2012}
to the G-MUSIC estimators with spatial smoothing in the asymptotic regime (\ref{eq:regime-M-N-L}). This is a difficult
task that is not within the scope of the present paper.
\end{remark}
Theorem \ref{theorem:spiked_eig} also allows to generalize immediately the results of \cite{vallet-loubaton-mestre-2014} concerning the consistency of the traditional estimates
$(\hat{\theta}_{k,N}^{(t)})_{k=1, \ldots, K}$ in the case $L=1$. In particular, while the traditional
estimate $\hat{\eta}_N^{(t)}$ of the pseudo-spectrum is not consistent, it is shown in
\cite{vallet-loubaton-mestre-2014} that if $L=1$, then
the arguments of its local minima $(\hat{\theta}_{k,N}^{(t)})_{k=1, \ldots, K}$ are consistent
and verify
\begin{equation}
\label{eq:M-consistency-MUSIC-DoA-estimates}
M \left( \hat{\theta}_{k,N}^{(t)} - \theta_k \right) \rightarrow 0 \; a.s.
\end{equation}
for each $k=1, \ldots, K$ if the separation condition is verified.
The reader can check that Theorem \ref{theorem:spiked_eig}
allows to generalize immediately this behaviour to the traditional DoA MUSIC
estimates with spatial smoothing in regime (\ref{eq:regime-M-N-L}). More precisely,
the following result holds.
\begin{theorem}
\label{th:consistency-MUSIC-DoA-estimates}
Under condition (\ref{eq:separation}), the DoA traditional MUSIC SS estimates $(\hat{\theta}_{k,N}^{(t)})_{k=1, \ldots, K}$
verify
\begin{equation}
\label{eq:M-consistency-GMUSIC-DoA-estimates}
M \left( \hat{\theta}_{k,N}^{(t)} - \theta_k \right) \rightarrow 0 \; a.s.
\end{equation}
for each $k=1, \ldots, K$.
\end{theorem}
\begin{remark}
\label{re:variances-identiques}
It is established in \cite{vallet-loubaton-mestre-2014} in the case $L = 1$ that if $\frac{{\bf S}_N{\bf S}_N^{*}}{N}$ converges towards a diagonal matrix, then $M^{3/2} \left( \hat{\theta}_{k,N}^{(t)} - \theta_k \right)$
has a Gaussian behaviour, and that the corresponding variance coincides with
the asymptotic variance of $M^{3/2} \left( \hat{\theta}_{k,N} - \theta_k \right)$. In particular, if
$L=1$, the asymptotic performance of MUSIC and G-MUSIC estimators coincide. It would be interesting to check whether this
result still holds true for the MUSIC and G-MUSIC estimators with spatial smoothing.
\end{remark}

Theorems \ref{thm:separation-spatial-smoothing} and \ref{th:consistency-GMUSIC-DoA-estimates}
as well as (\ref{eq:M-consistency-MUSIC-DoA-estimates}) assume that the DoAs
$(\theta_k)_{k=1, \ldots, K}$ are fixed parameters, i.e. do not scale with $M$.
Therefore, the ratio
$$
\frac{\min_{k \neq l} |\theta_k - \theta_l|}{(2 \pi)/M}
$$
converges towards $+\infty$. In practice, this context is thus able to
model practical situations in which $\sup_{k \neq l} |\theta_k - \theta_l|$
is significantly larger than the aperture of the array. In the case $L=1$,
\cite{vallet-loubaton-mestre-2014} also addressed the case where the DoA's $(\theta_{k,N})_{k=1, \ldots, K}$ depend on $N,M$ and verify $\theta_{k,N} - \theta_{l,N} = \mathcal{O}(\frac{1}{M})$. This context
allows to capture practical situations in which the DoA's are spaced of the order of a beamwidth.
In order to simplify the calculations, \cite{vallet-loubaton-mestre-2014} considered the case
$K=2$, $\theta_{2,N} = \theta_{1,N} + \frac{\alpha}{N}$ and where matrix $\frac{{\bf S}_N{\bf S}_N^{*}}{N} \rightarrow {\bf I}_2$. However,
the results can be generalized easily to more general situations. It is shown
in \cite{vallet-loubaton-mestre-2014} that the G-MUSIC estimates still verifiy
(\ref{eq:M-consistency-GMUSIC-DoA-estimates}), but that, in general,
$M \left( \hat{\theta}_{k,N}^{(t)} - \theta_k \right)$ does not converge towards $0$.
The results of \cite{vallet-loubaton-mestre-2014} can be generalized immediately
to the context of G-MUSIC estimators with spatial smoothing in regime (\ref{eq:regime-M-N-L}).
For this, we have to assume that $\theta_{2,N} = \theta_{1,N} + \frac{\kappa}{M}$
(in  \cite{vallet-loubaton-mestre-2014}, $M$ and $N$ are of the same order of magnitude so
that the assumptions $\theta_{2,N} = \theta_{1,N} + \frac{\alpha}{N}$ and $\theta_{2,N} = \theta_{1,N} + \frac{\kappa}{M}$ are equivalent), and to follow the arguments of section 4 in
\cite{vallet-loubaton-mestre-2014}. The conclusion of this discussion is the following
Theorem.
\begin{theorem}
\label{th:closely-spaced-DoA}
Assume $K=2$, $\theta_{2,N} = \theta_{1,N} + \frac{\kappa}{M}$, and that
$\frac{{\bf S}_N{\bf S}_N^{*}}{N} \rightarrow {\bf I}_2$. If the separation condition
\begin{equation}
\label{eq:separation-condition-closely-spaced}
1 - | \sinc \kappa/2 | > \sigma^{2} c_*
\end{equation}
holds, then the G-MUSIC SS estimates $(\hat{\theta}_{k,N})_{k=1,2}$ defined by
\begin{equation}
\label{eq:def-theta-GMUSIC-closely-spaced}
	\hat{\theta}_{k,N} = \argmin_{\theta \in \Ical_{k,N}} \left| \hat{\eta}_N(\theta) \right|,
\end{equation}
where $\Ical_{k,N} = [\theta_{k,N} - \frac{\kappa - \epsilon}{2N}, \theta_{k,N} + \frac{\kappa - \epsilon}{2N}]$ for $\epsilon$ small enough, verify
\begin{equation}
\label{eq:M-consistency-GMUSIC-closely-space-DoA-estimates}
M \left( \hat{\theta}_{k,N} - \theta_{k,N} \right) \rightarrow 0 \; a.s.
\end{equation}
In general, the traditional MUSIC SS estimates defined by (\ref{eq:def-theta-GMUSIC-closely-spaced})
when the G-MUSIC estimate $\hat{\eta}_N(\theta)$ is replaced by the traditional spectrum estimate
$\hat{\eta}_N^{(t)}(\theta)$ are such that
$M \left( \hat{\theta}_{k,N}^{(t)} - \theta_{k,N} \right)$ does not converge towards $0$.
\end{theorem}

	\section{Numerical examples}
	\label{sec:simulation}
	
In this section, we provide numerical simulations illustrating the results given
in the previous sections. We first consider 2 closely spaced sources
with DoAs $\theta_1 = 0$ and $\theta_2 = \frac{\pi}{2M}$, and we assume that $M=160$ and
$N=20$. The $2 \times N$ signal matrix is obtained by normalizing a realization of a random matrix with
$\Ncal_{\Cbb}(0,1)$ i.i.d. entries in such a way that the 2 source signals have power 1. The signal
to noise ratio is thus equal to $\mathrm{SNR} = 1/\sigma^2$. Table \ref{table:L_SNR} provides the minimum value of SNR for which the separation condition, in its finite length version (i.e. when the limits $(\lambda_k)_{k=1, \ldots, K}$ and
$c_*$ are replaced by $(\lambda_{k,N})_{k=1, \ldots, K}$ and
$c_N$ respectively)  holds, i.e.
$$
(\sigma^{2})^{-1} = \frac{1}{\lambda_{K,N}}  \sqrt{(M-L+1)/NL}
$$
It is seen that the minimal SNR first decreases but that
it increases if $L$ is large enough. This confirms the discussion of
the previous section on the effect of $L$ on the separation condition.

\begin{table}[h!]
\centering
\resizebox{1\columnwidth}{!}{%
  \begin{tabular}{|c|c|c|c|c|c|c|c|c|}
    \hline
    \bf{L} &2    &4    &8   &16   &32   &64   &96
     &128 \\
     \hline
    \bf{SNR}  &33.46   &30.30   &27.46   &25.31   &24.70   &28.25   &36.11
      &51.52 \\
    \hline
   \end{tabular}
}
\caption{Minimum value of SNR for separation condition }
\label{table:L_SNR}
\end{table} 	

\begin{figure}[h!]
\centering
	\includegraphics[scale=0.5]{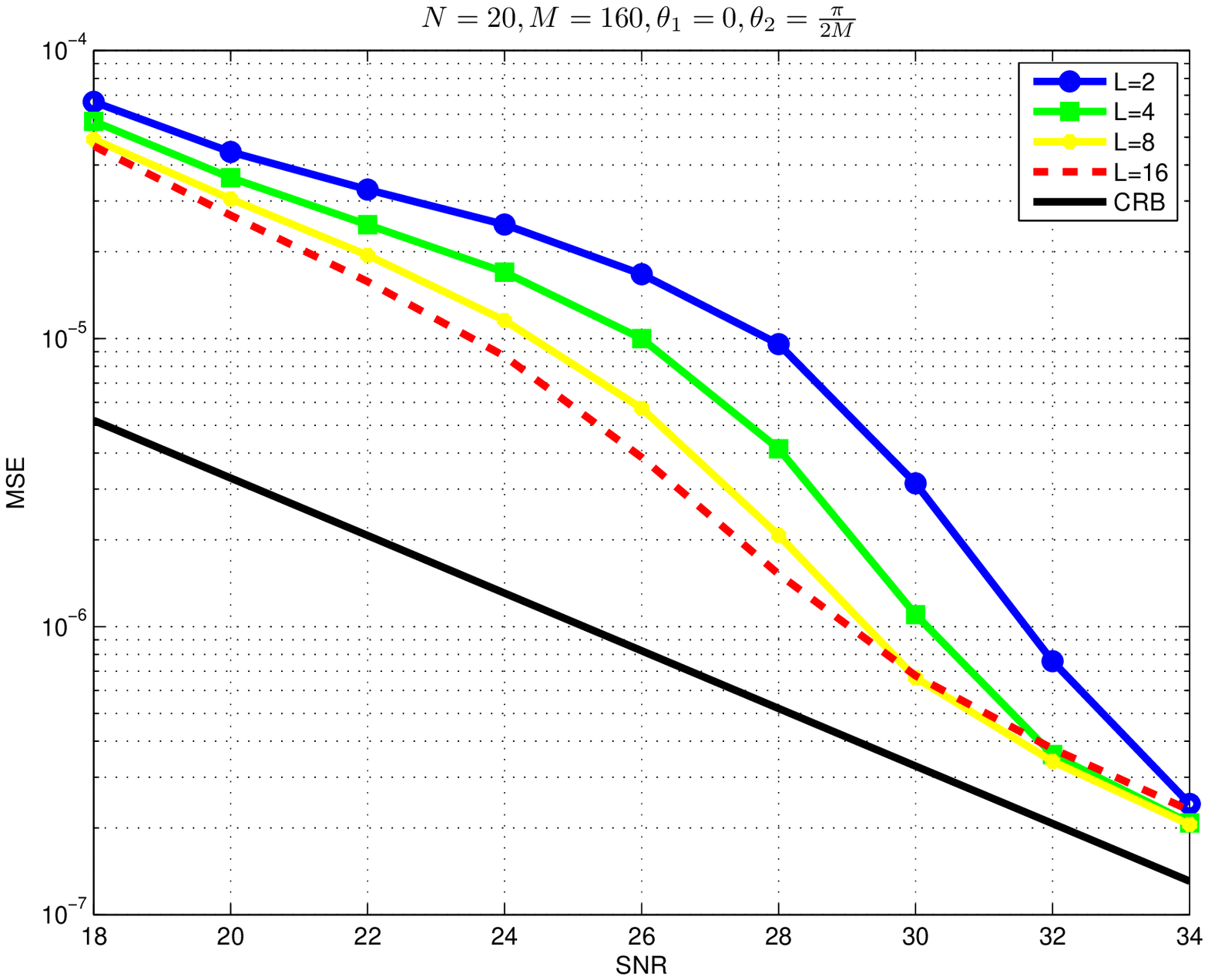}
	\caption{Empirical MSE of G-MUSIC SS estimator $\hat{\theta}_1$ versus SNR}
    \label{figure:mse_ts_L_1_16}
\end{figure}

\begin{figure}[h!]
\centering
    \includegraphics[scale=0.5]{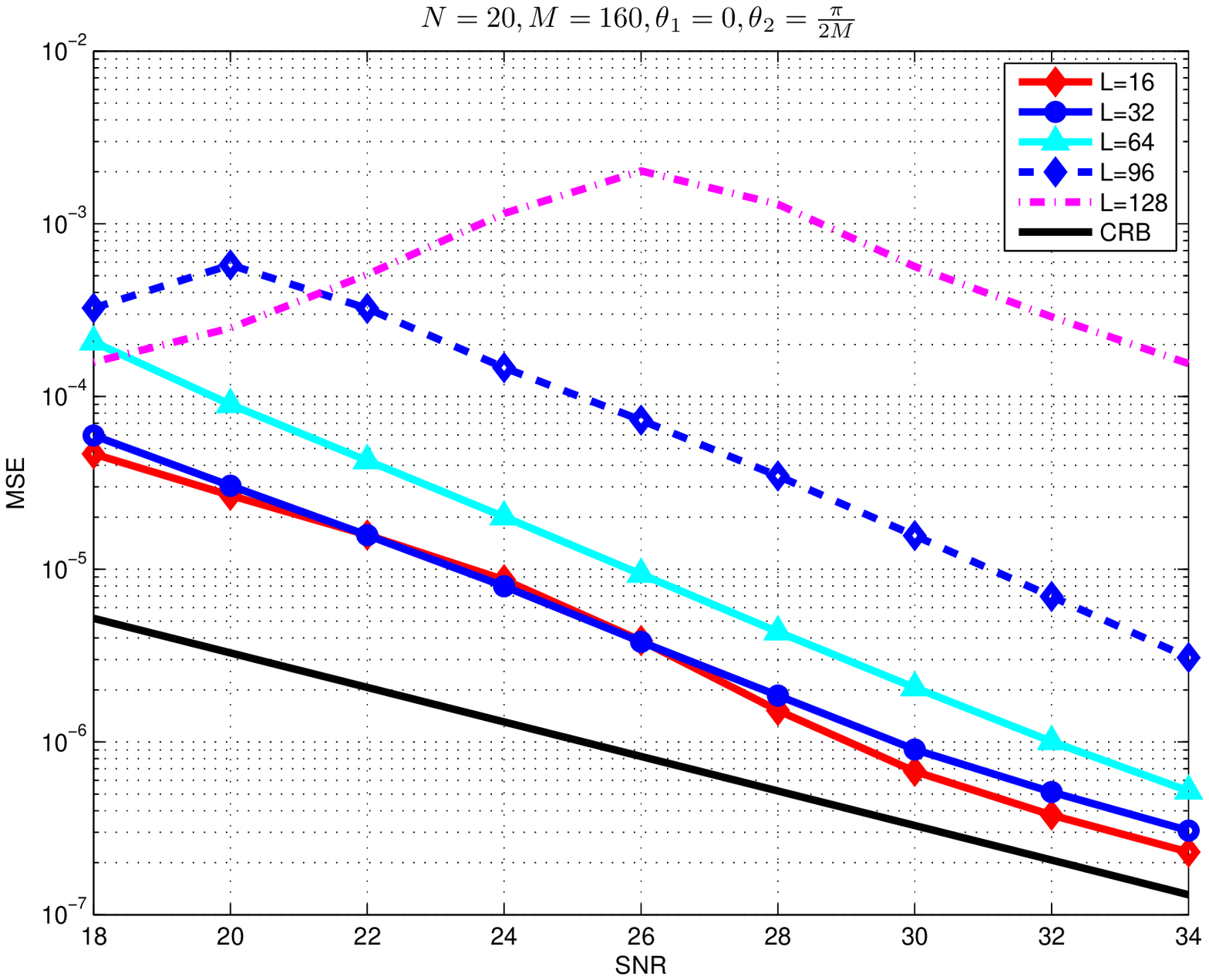}
	\caption{Empirical MSE of G-MUSIC SS estimator $\hat{\theta}_1$ versus SNR}
	\label{figure:mse_ts_L_16_128}
\end{figure}

In figure \ref{figure:mse_ts_L_1_16}, we represent
the mean-square errors of  the G-MUSIC SS estimator $\hat{\theta}_1$ for $L=2, 4, 8, 16$ versus
SNR. The corresponding Cramer-Rao bounds is also represented. As expected, it is seen that the performance tends to increase with $L$ until
$L=16$. In figure \ref{figure:mse_ts_L_16_128}, $L$ is equal to 16, 32, 64, 96,
128.

For $L=32$, it is seen that the MSE tends to degrade at high SNR
w.r.t. $L=16$, while the performance severely degrades for larger values of $L$.

\begin{figure}[h!]	
\centering
	\includegraphics[scale=0.5]{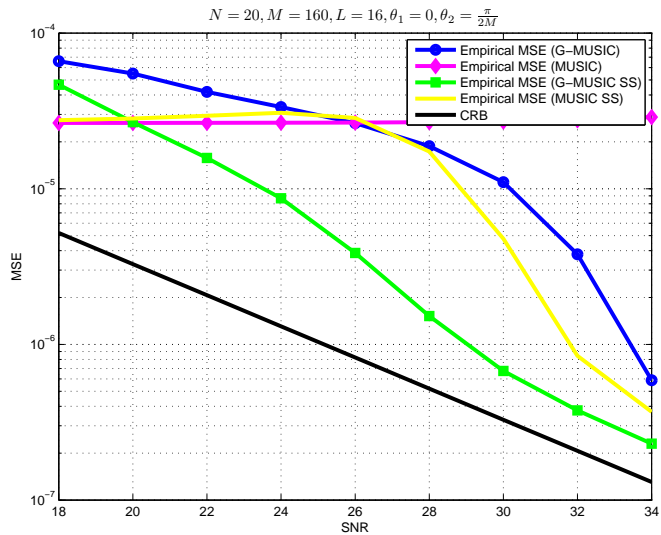}
	\caption{Empirical MSE of different estimators of $\theta_1$ when L=16}
	\label{figure:mse_L_16}
\end{figure}

In Figure \ref{figure:mse_L_16}, parameter $L$ is equal to $16$.
We compare the performance of G-MUSIC SS with the standard MUSIC method with spatial
smoothing. We also represent the MSE provided by G-MUSIC and MUSIC for $L=1$.
The standard unsmoothed MUSIC method of course completely fails, while the use of the
G-MUSIC SS provides a clear improvement of the performance w.r.t. MUSIC SS and unsmoothed
G-MUSIC.

We finally consider the case $L=128$, and compare as above G-MUSIC SS, MUSIC SS,
unsmoothed G-MUSIC and unsmoothed MUSIC. G-MUSIC SS completely fails because
$L$ and $M$ are of the same order of magnitude.
Theorem \ref{thm:separation-spatial-smoothing} is thus no more valid, and the pseudo-spectrum estimate
is not consistent.

\begin{figure}[h!]
\centering
	\includegraphics[scale=0.5]{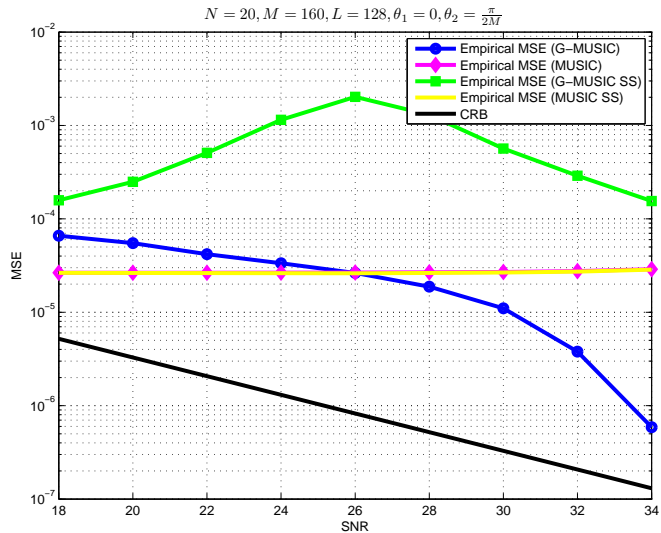}
	\caption{Empirical MSE of different estimators of $\theta_1$ when L=128}
	\label{figure:mse_L_128}
\end{figure}	

\begin{figure}[h!]
\centering
	\includegraphics[scale=0.5]{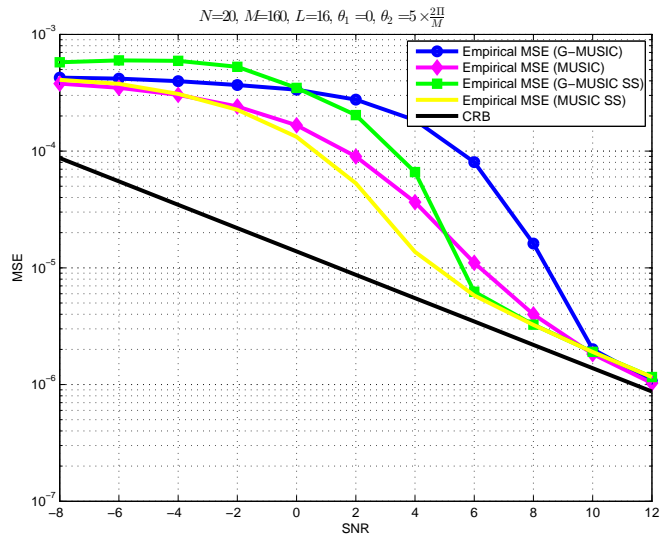}
	\caption{Empirical MSE of different estimators of $\theta_1$ when L=16 and widely spaced DoAs}
	\label{figure:mse_L_16_anglelointain}
\end{figure}

We now consider 2 widely spaced sources with DoAs $\theta_1 = 0$ and $\theta_2 = 5 \frac{2\pi}{M}$,
and keep the same parameters as above. We consider the case $L=16$, and represent
in Fig. \ref{figure:mse_L_16_anglelointain} the performance of MUSIC, G-MUSIC, MUSIC-SS, and
G-MUSIC-SS. It is first observed that, in contrast with the case of closely spaced DoAs, MUSIC-SS and G-MUSIC-SS have the same performance when the SNR is above the threshold 6 dB. This is in accordance
with Theorem \ref{th:consistency-MUSIC-DoA-estimates}, and tends to indicate that, as in the
case $L=1$, if $\frac{{\bf S}_N {\bf S}_N^{*}}{N}$ converges towards a diagonal matrix,
then the asymptotic performance of G-MUSIC-SS and MUSIC-SS coincide (see Remark \ref{re:variances-identiques}). The comparison between
the methods with and without spatial smoothing also confirm that the use of spatial smoothing schemes
allow to improve the performance.
	
\section{Conclusion}

In this paper, we have addressed the behaviour of subspace DoA estimators
based on spatial smoothing in asymptotic regimes where $M$ and $NL$ converge towards
$\infty$ at the same rate. For this, we have evaluated the behaviour of the largest singular values and
corresponding singular vectors of large random matrices defined as additive low rank
perturbations of certain random block-Hankel matrices, and established that
they behave as if the entries of the block-Hankel matrices were i.i.d. Starting from
this result, we have shown that it is possible to generalize the G-estimators introduced in
\cite{hachem2012subspace}, and have deduced from \cite{vallet-loubaton-mestre-2014} their properties.

\appendices
\section{Insights on the proof of (\ref{eq:convergence-lambda-spike}) and (\ref{eq:converge-lambda-bulk}).}
We first recall that \cite{benaych2012singular} established
that for $1 \leq k \leq K$, if $\hat{\lambda}_{k,N}$ does not converge towards
a limit strictly greater than $x_*^{+}$, then $\hat{\lambda}_{k,N}$ converges towards
$x_*^{+}$. We have therefore to evaluate the behaviour of the eigenvalues
of ${\bf X}_N {\bf X}_N^{*}$ that are greater than $x_*^{+}+\epsilon$
for some $\epsilon > 0$.

If ${\bf C}$ represents a $P \times Q$ matrix,
we denote by $\underline{{\bf C}}$ the $(P+Q) \times (P+Q)$ hermitian matrix defined by
\begin{equation}
\label{eq:def-underline}
\underline{{\bf C}} =\left[\begin{array}{cc}
  0 & {\bf C} \\
  {\bf C}^* & 0
\end{array}\right].
\end{equation}

Then, the non zero eigenvalues of $\underline{{\bf C}}$ coincide with the (positive and
negative) square roots of the non zero eigenvalues of matrix ${\bf C} {\bf C}^{*}$,
and the corresponding eigenvectors are the $(P+Q)$--dimensional vectors $({\bf a}_k^{T},  \pm {\bf b}_k^{T})^{T}$ where $({\bf a}_k, {\bf b}_k)$ represent the pairs of left and right singular vectors of ${\bf C}$.
Therefore, $\lambda > x_*^{+} + \epsilon$ is eigenvalue of ${\bf X}_N {\bf X}_N^{*}$
if and only if $\sqrt{\lambda} > (x_*^{+} + \epsilon)^{1/2}$ is eigenvalue of
matrix $\underline{{\bf X}}_N$. We consider the singular value decomposition
$$
{\bf B}_N = {\bf U}_N {\bs \Lambda}_N^{1/2} \tilde{{\bf U}}_N^{*}
$$
of matrix ${\bf B}_N$ and express $\underline{{\bf X}}_N$ as
$$
\underline{{\bf X}}_N = \left[\begin{array}{cc}
0 & {\bf Z}_N \\
{\bf Z}_N^*& 0
\end{array}\right ] + \left[\begin{array}{cc}
0 &\U_N {\bs \Lambda}_N^{1/2} \tilde{{\bf U}}_N^* \\
\tilde{{\bf U}}_N {\bs \Lambda}_N^{1/2} \U_N^*& 0
\end{array}\right ]
$$
which can be written as
$$
\left[\begin{array}{cc}
0 & {\bf Z}_N \\
{\bf Z}_N* & 0
\end{array}\right ] +
\underbrace{\left[\begin{array}{cc}
\U_N & 0  \\
0 & \tilde{{\bf U}}_N {\bs \Lambda}_N^{1/2}
\end{array}\right ]
}_{{\bf D}} {\bf J}
\underbrace{\left[\begin{array}{cc}
\U_N^* & 0  \\
0 & {\bs \Lambda}_N^{1/2} \tilde{{\bf U}}_N^*
\end{array}\right ]
}_{{\bf D}^*}
$$
where ${\bf J}$ is defined by
$$
{\bf J} = \left( \begin{array}{cc} 0 & {\bf I}_K \\
                                   {\bf I}_K & 0 \end{array} \right)
$$
Consider $x > (x_*^{+} + \epsilon)^{1/2}$. Then, $x$ is not a singular value of
${\bf Z}_N {\bf Z}_N^{*}$, and therefore, not an eigenvalue of $\underline{{\bf Z}}_N$. Therefore,
it holds that
\begin{eqnarray*}
\det(\underline{{\bf X}}_N -x\I) \hspace{-0.2cm} &=& \det(\underline{{\bf Z}}_N - x\I + {\bf D} \J {\bf D}^*)  \\
                                     &=& \det(\underline{{\bf Z}}_N - x\I)\det(\I + (\underline{{\bf Z}}_N - x\I)^{-1} {\bf D} \J {\bf D}^*) \\
                                     &=& \det(\underline{{\bf Z}}_N - x\I) \det(\I_{2K} + \J {\bf D}^*(\underline{{\bf Z}}_N - x\I)^{-1} {\bf D})
\end{eqnarray*}
after noticing $\J=\J^{-1}$. For $w \in \mathbb{C} - [-(x_*^{+} + \epsilon)^{1/2}, (x_*^{+} + \epsilon)^{1/2}]$, we denote by ${\bf S}_N(w)$ the $2K \times 2K$ matrix defined by
$$
{\bf S}_N(w) = \I_{2K} + \J {\bf D}^*(\underline{{\bf Z}}_N - w\I)^{-1} {\bf D}
$$
Using the identity
$$
(\underline{{\bf Z}}_N-w\I)^{-1}=
\left[\begin{array}{cc}
w {\bf Q}_N(w^2) & {\bf Q}_N(w^2) {\bf Z}_N\\
{\bf Z}_N^* {\bf Q}_N(w^2) & w\tilde{{\bf Q}}_N(w^2)
\end{array}\right]
$$
we obtain immediately that
\begin{multline*}
\left({\bf S}_N(w)\right)_{1,1} = \I_K + {\bs \Lambda}_N^{1/2} \tilde{{\bf U}}_N^* {\bf Z}_N^*{\bf Q}_N(w^2)\U_N \\
\left({\bf S}_N(w)\right)_{1,2} = w {\bs \Lambda}_N^{1/2} \U_N^*\widetilde{{\bf Q}}_N(w^2) \tilde{{\bf U}}_N {\bs \Lambda}_N^{1/2} \\
\left({\bf S}_N(w)\right)_{2,1} = w\U_N^*{\bf Q}_N(w^2)\U_N \\
\left({\bf S}_N(w)\right)_{2,2} = \I_K + \U_N^*{\bf Q}_N(w^2) {\bf Z}_N \tilde{{\bf U}}_N {\bs \Lambda}_N^{1/2}
\end{multline*}
Item (iii) of Proposition \ref{prop:eigenvalues-ZZ*} implies that the elements of ${\bf S}_N(w)$
converge almost surely, uniformly on the compact subsets of $\mathbb{C} - [-(x_*^{+} + \epsilon)^{1/2}, (x_*^{+} + \epsilon)^{1/2}]$ towards the elements of matrix ${\bf S}_*(w)$ defined by
$$
{\bf S}_*(w) =
\left[\begin{array}{cc}
\I_K & w\widetilde{m}_*(w^2){\bs \Lambda} \\
wm(w^2)\I_K & \I_K
\end{array}\right]
$$
It is easy to check that $\det({\bf S}_N(w))$ and $\det({\bf S}_*(w))$ are functions of $w^{2}$.
We define functions $s_N$ and $s_*$ on $\mathbb{C} - [0, x_*^{+} + \epsilon]$ by
$s_N(w^{2}) = \det({\bf S}_N(w))$ and $s_*(w^{2}) = \det({\bf S}_*(w))$.
It is clear that almost surely, $s_N(z) \rightarrow s_*(z)$ uniformly on the compact subsets of
$\mathbb{C} - [0, (x_*^{+} + \epsilon)]$. Therefore, in order to precise the behaviour of the
eigenvalues of ${\bf X}_N {\bf X}_N^{*}$ that are greater than $x_*^{+} + \epsilon$ (i.e. the solutions of the equation $s_N(x) = 0$ greater than $x_*^{+} + \epsilon$), it is first useful to characterize the solutions of the equation $s_*(x) = 0$.
The equation $s_*(x) = 0$ is equivalent to
$$
\Pi_{k=1}^{K} \left( 1 - \lambda_k x m_*(x) \tilde{m}_*(x) \right) = 0
$$
or equivalently to
$$
w_*(x) = \lambda_k
$$
for $k=1, \ldots, K$. Using the properties of function $w_*$, we obtain immediately that
if $\epsilon < \rho_s - x_*^{+} = \phi_*(\lambda_s) - x_*^{+}$, then the solutions of $s_*(x) = 0$ that are greater than $x_*^{+} + \epsilon$ coincide with the $(\rho_k)_{k=1, \ldots, s}$ defined by $\rho_k = \phi_*(\lambda_k)$ for $k=1, \ldots, s$. Using this, it can be proved using appropriate arguments that, almost surely, for $N$ large enough, then the $s$ greatest
eigenvalues  $(\lambda_{k,N})_{k=1, \ldots, s}$ of ${\bf X}_N {\bf X}_N^{*}$ are greater than
$x_*^{+} + \epsilon$, and that $\lambda_{k,N} \rightarrow \rho_k$ for $k=1, \ldots, s$ \footnote{The arguments used in \cite{benaych2012singular} require the uniform convergence of $s_N$ towards $s_*$ on the set $\mathrm{Re}(z) > \x_*^{+} + \epsilon$, a property that is not established in Proposition \ref{prop:eigenvalues-ZZ*}. However, the proof of the
contuinity lemma 2.1 in \cite{benaych2011square} can be simplified, and only needs the uniform convergence
on compact sets.}.

\section{Proof of Proposition \ref{prop:eigenvalues-ZZ*}.}
The proof of Proposition \ref{prop:eigenvalues-ZZ*}
is based on the results of \cite{loubaton-bloc-hankel}. In order to explain this, we denote by ${\bf W}_N$ the $NL \times (M-L+1)$ matrix defined by
$$
{\bf W}_N = \frac{1}{\sqrt{c_N}}  \, {\bf Z}_N^{*}
$$
The variance of the entries of ${\bf W}_N$ is equal to $\frac{\sigma^{2}}{M-L+1}$. Therefore, matrix ${\bf W}_N$ is similar to the matrices studied in \cite{loubaton-bloc-hankel}
except that the integers $(M,N)$ in \cite{loubaton-bloc-hankel} should be exchanged by
$(N,M-L+1)$. In particular, after this replacement, it is clear that the asymptotic regime
(\ref{eq:regime-M-N-L}) coincides with the regime in \cite{loubaton-bloc-hankel}. In order to
recall the results of \cite{loubaton-bloc-hankel}, we denote by $t_N(z)$, $\tilde{t}_N(z)$,
$t_*(z)$ and $\tilde{t}_*(z)$ the Stieltjes transforms of the Marcenko-Pastur distributions
of parameters $(\sigma^{2}, c_N^{-1})$, $(\sigma^{2} c_N^{-1}, c_N)$,
$(\sigma^{2}, c_*^{-1})$ and $(\sigma^{2} c_*^{-1}, c_*)$. Moreover, ${\bf Q}_{N,W}(z)$
and $\tilde{{\bf Q}}_{N,W}(z)$ represent
the resolvents of matrices ${\bf W}_N {\bf W}_N^{*}$ and ${\bf W}_N^{*} {\bf W}_N$ respectively.
It is shown in \cite{loubaton-bloc-hankel} (see Section 6) that the eigenvalue distribution of
${\bf W}_N {\bf W}_N^{*}$ converges almost surely towards $\mu_{\sigma^{2}, c_*^{-1}}$,
a statement equivalent to
$$
\frac{1}{NL} \mathrm{Tr}({\bf Q}_{N,W}(z)) - t_*(z) \rightarrow 0 \; a.s.
$$
or to
$$
\frac{1}{M-L+1} \mathrm{Tr}(\tilde{{\bf Q}}_{N,W}(z)) - \tilde{t}_*(z) \rightarrow 0 \; a.s.
$$
for each $z \in \mathbb{C}^{+}$. As we have
\begin{equation}
\label{eq:expre-Z*Z}
{\bf Z}_N^{*} {\bf Z}_N = c_N  {\bf W}_N {\bf W}_N^{*}
\end{equation}
it holds that the resolvent $\tilde{{\bf Q}}_N(z)$ of ${\bf Z}_N^{*} {\bf Z}_N$ is equal to
$$
\tilde{\Q}_N(z) = c_N^{-1} {\Q}_{N,W}(z c_N^{-1})
$$
As $c_N \rightarrow c_*$, it is clear $\frac{1}{NL} \mathrm{Tr}(\tilde{{\bf Q}}_N(z))$
behaves as $\frac{1}{c_*} \frac{1}{NL} \mathrm{Tr}( {\bf Q}_{N,W}(z c_{*}^{-1})$.
Similarly, $\frac{1}{M-L+1} \mathrm{Tr}({\bf Q}_N(z))$ behaves as
$\frac{1}{c_*} \frac{1}{M-L+1} \mathrm{Tr}(\tilde{{\bf Q}}_{N,W}(z c_{*}^{-1}))$.
Therefore, for each $z \in \mathbb{C}^{+}$, it holds that
$$
\frac{1}{NL} \mathrm{Tr}(\tilde{\Q}_N(z)) - c_*^{-1} \, t_*(z c_{*}^{-1}) \rightarrow 0 \; a.s.
$$
and that
$$
\frac{1}{M-L+1} \mathrm{Tr}({\bf Q}_N(z)) - c_*^{-1} \, \tilde{t}_*(z c_{*}^{-1}) \rightarrow 0 \; a.s.
$$
Using Equations (\ref{eq:fundamental_equation_t-ttilde-1}, \ref{eq:fundamental_equation_t-ttilde-2}),
it is easy to verify that $m_*(z) = c_*^{-1} \, \tilde{t}_*(z c_{*}^{-1})$ and
$\tilde{m}_*(z) = c_*^{-1} \, t_*(z c_{*}^{-1})$. This establishes
(\ref{eq:convergence-MP-ZZ*}) and the convergence of the eigenvalue distribution
of ${\bf Z}_N {\bf Z}_N^{*}$ towards $\mu_{\sigma^{2},c_*}$.

Asymptotic regime (\ref{eq:regime-M-N-L}) implies that $L = \mathcal{O}(M^{\alpha}) =
\mathcal{O}((M-L+1)^{\alpha})$ where
$\alpha < 2/3$. Therefore, \cite{loubaton-bloc-hankel} implies that for each
$\epsilon > 0$,
almost surely, for $N$ large enough, the eigenvalues of ${\bf W}_N {\bf W}_N^{*}$
are located in $[\sigma^{2}(1 - \sqrt{c_*^{-1}})^{2} - \epsilon, \sigma^{2}(1 + \sqrt{c_*^{-1}})^{2}]
\cup \{ 0 \} \mathbb{1}{(c_*^{-1} > 1)}$. (\ref{eq:expre-Z*Z}) and the convergence of $c_N$ towards
$c_*$ lead immediately to item (ii) of Proposition \ref{prop:eigenvalues-ZZ*}.

Using the same arguments as above, (\ref{eq:convergence-forme-quadratique-tildeQ}) appears as
a consequence of
\begin{equation}
\label{eq:convergence-forme-quadratique-QW}
\tilde{{\bf a}}_N^{*} \left( {\bf Q}_{N,W}(z) - t_{*}(z) {\bf I} \right) \tilde{{\bf b}}_N \rightarrow 0 \; a.s.
\end{equation}
While (\ref{eq:convergence-forme-quadratique-QW}) does not appear explicitely in
\cite{loubaton-bloc-hankel}, it can be deduced rather easily from the various
intermediate results proved in \cite{loubaton-bloc-hankel}. For this,
we first remark that
\begin{equation*}
\tilde{{\bf a}}_N^{*} \left( {\bf Q}_{N,W}(z) - t_{*}(z) {\bf I} \right) \tilde{{\bf b}}_N =
\tilde{{\bf a}}_N^{*} \left( {\bf Q}_{N,W}(z) - \mathbb{E}( {\bf Q}_{N,W}(z)) \right) \tilde{{\bf b}}_N
+\tilde{{\bf a}}_N^{*} \left( \mathbb{E}({\bf Q}_{N,W}(z)) - t_{*}(z) {\bf I} \right) \tilde{{\bf b}}_N
\end{equation*}
and establish that the 2 terms at the right hand side of the above equation converge towards $0$.
In order to simplify the notations,  we denote by $\xi$ the first term. The almost sure convergence of $\xi$ towards $0$ follows from the Poincar\'e-Nash inequality (see e.g. Proposition 2 of \cite{loubaton-bloc-hankel}). Exchanging
$(M,N)$ by $(N,M-L+1)$ in Proposition 6 of \cite{loubaton-bloc-hankel}, we obtain immediately
that $\mathbb{E}|\xi|^{2} = \mathcal{O}(\frac{L}{M-L+1}) = \mathcal{O}(\frac{L}{M})$. As $L/M \rightarrow 0$, this implies that $\xi$ converges in probability towards $0$. In order to prove the almost sure
convergence, we briefly justify that for each $n$, it holds that
\begin{equation}
\label{eq:moment-xi}
\mathbb{E}|\xi|^{2n} = \mathcal{O}\left( (L/M)^{n} \right)
\end{equation}
(\ref{eq:moment-xi}) can be established by induction on $n$. As mentioned above, (\ref{eq:moment-xi}) is verified for $n=1$. We now assume that it holds until integer $n-1$, and prove (\ref{eq:moment-xi}).
For this, we use the obvious relation:
$$
\mathbb{E}|\xi|^{2n} = \left( \mathbb{E}|\xi|^{n} \right)^{2} + \mathrm{Var}(\xi^{n})
$$
Using the Poincar\'e-Nash inequality as in the proof of Proposition 6 of \cite{loubaton-bloc-hankel},
we obtain easily that
$$
\mathrm{Var}(\xi^{n}) \leq C \,\frac{L}{M} \mathbb{E}(|\xi|^{2n-2})
$$
where $C$ is a constant that depends on $z$ but not on the dimensions $L,M,N$. As (\ref{eq:moment-xi}) is assumed to hold until
integer $n-1$, this implies that $\mathrm{Var}(\xi^{n}) = \mathcal{O}\left( (L/M)^n \right)$.
The Schwartz inequality leads immediately to
$$
\left( \mathbb{E}|\xi|^{n} \right)^{2} \leq \mathbb{E}(|\xi|^{2}) \, \mathbb{E}(|\xi|^{2n-2})
$$
which is a $\mathcal{O}\left( (L/M)^n \right)$ term. This establishes (\ref{eq:moment-xi}).
As $L = \mathcal{O}(M^{\alpha})$ with $\alpha < 2/3$, it is clear that $(L/M)^{3}$
verifies
$$
(L/M)^{3} = \frac{1}{M^{1+2-3\alpha}}
$$
Therefore, (\ref{eq:moment-xi}) for $n=3$ leads to
$$
\mathbb{E}\left(|\xi|^{6}\right) = \mathcal{O}\left(  \frac{1}{M^{1+2-3\alpha}} \right)
$$
As $2 - 3 \alpha > 0$, the use of the Markov inequality and of the Borel-Cantelli lemma imply that
$\xi$ converges towards $0$ almost surely as expected.

It remains to justify that $\tilde{{\bf a}}_N^{*} \left( \mathbb{E}({\bf Q}_{N,W}(z)) - t_{*}(z) {\bf I} \right) \tilde{{\bf b}}_N$ converges towards $0$. Although it is not stated explicitely in \cite{loubaton-bloc-hankel}, it can immediately deduced from Eq. (5.3) in Proposition 8, as well
as on Corollary 1, Theorem 2, and formula (7.3). \\

(\ref{eq:convergence-forme-quadratique-Q}) is equivalent to
\begin{equation}
\label{eq:convergence-forme-quadratique-tildeQW}
{\bf a}_N^{*} \left( \tilde{{\bf Q}}_{N,W}(z) - \tilde{t}_{*}(z) {\bf I} \right) {\bf b}_N \rightarrow 0 \; a.s.
\end{equation}
It can be proved as above that
$$
{\bf a}_N^{*} \left( \tilde{{\bf Q}}_{N,W}(z) - \mathbb{E}(\tilde{{\bf Q}}_{N,W}(z)) \right) {\bf b}_N \rightarrow 0 \; a.s.
$$
and establish that
\begin{equation}
\label{eq:convergence-esperance-forme-quadratique-tildeQW}
{\bf a}_N^{*} \left( \mathbb{E}(\tilde{{\bf Q}}_{N,W}(z)) - \tilde{t}_{*}(z) {\bf I} \right) {\bf b}_N \rightarrow 0
\end{equation}
for each $z \in \mathbb{C}^{+}$. The behaviour of matrix $\mathbb{E}(\tilde{{\bf Q}}_{N,W}(z))$ is not studied in \cite{loubaton-bloc-hankel}.
However, it can be evaluated using the results of \cite{loubaton-bloc-hankel}. For this, we first simplify the notations and denote by $\W, \tilde{\W}, \Q, \tilde{\Q}$ the matrices $\W_N, \tilde{\W}_N, \Q_{N,W}(z)$, and $ \tilde{\Q}_{N,W}(z)$. Moreover, $\Q$ is a $NL\times NL$ block matrix, so that we denote by $\Q^{n_1,n_2}_{i_1,i_2}$  its entry $(i_1 + (n_1-1)L,i_2 + (n_2-1)L)$.

As in \cite{loubaton-bloc-hankel}, we denote by $\tau^{(N)}(.)$ and $\mathcal{T}^{(N)}_{M-L+1,L}(.)$ the operators defined by
\begin{eqnarray*}
\tau^{(N)}(\Q)(i) = \frac{1}{NL} \Tr(\Q (\I\otimes \J^i_L)) \\
\mathcal{T}^{(N)}_{M-L+1,L}(\Q) = \sum^{L-1}_{i=-(L-1)} \tau^{(N)}(\Q)(i) \J^{*i}_{M-L+1}
\end{eqnarray*}
where $\J_L$ is the $L\times L$ "shift" matrix defined by $(\J_L)_{i,j}= \delta(j-i=1)$.

Replacing integers $(M,N)$ by integers $(N, M-L+1)$ in Equation (4.6) of \cite{loubaton-bloc-hankel},
we obtain that
\begin{eqnarray*}
\mathbb{E}\left[ (\Q\W)^{n_1}_{i_1,k} (\W^*)^{n_2}_{j,i_2} \right] = \frac{\sigma^2}{M-L+1} \mathbb{E} \left(\Q^{n_1,n_2}_{i_1,i_2-(k-j)}\right) \mathbb{1}_{1\leq i_2-(k-j)\leq L}  \\
- \frac{\sigma^2}{c_N} \sum^{L-1}_{i= -(L-1)} \mathbb{1}_{1\leq k-i\leq M-L+1} \mathbb{E} \left[\tau^{(N)}(\Q)(i) (\Q\W)^{n_1}_{i_1,k-i} (\W^*)^{n_2}_{j,n_2} \right]
\end{eqnarray*}
with  $1 \leq j,k \leq M-L+1$, $1\leq n_1,n_2\leq N$, $1\leq i_1,i_2\leq L$.

Setting $u=k-i$, the second term of the righthandside of the above equation can also be written as
$$
\frac{\sigma^2}{c_N} \sum^{M-L+1}_{u=1} \mathbb{E} \left[\tau^{(N)}(\Q)(k-u) \mathbb{1}_{-(L-1)\leq k-u\leq L-1} (\Q\W)^{n_1}_{i_1,u} (\W^*)^{n_2}_{j,n_2}\right]
$$
Now setting $n=n_1=n_2,i=i_1=i_2$, and summing over $n$ and $i$, we obtain
\begin{eqnarray*}
\mathbb{E} \left(\W^* \Q \W \right)_{j,k} = \frac{\sigma^2}{c_N} \tau^{(N)}(\mathbb{E}(\Q))(k-j)\mathbb{1}_{-(L-1)\leq k-j\leq L-1} \\
- \frac{\sigma^2}{c_N} \mathbb{E}\left( \sum^{M-L+1}_{u=1} \tau^{(N)}(\Q)(k-u)\mathbb{1}_{-(L-1)\leq k-u\leq L-1} (\W^*\Q\W)_{j,u}\right)
\end{eqnarray*}
and using that $\tau^{(N)}(\Q)(k-u)\mathbb{1}_{-(L-1)\leq k-u\leq L-1} = \left(\mathcal{T}^{(N)}_{M-L+1,L}(\Q)\right)_{k,u}$, we get that
\begin{equation*}
\mathbb{E} \left(\W^* \Q\W \right)_{j,k} = \frac{\sigma^2}{c_N} \left(\mathcal{T}^{(N)}_{M-L+1,L}(\mathbb{E}(\Q))\right)_{k,j}
- \frac{\sigma^2}{c_N} \mathbb{E}\left( \mathcal{T}^{(N)}_{M-L+1,L}(\Q) \W^T \Q^T \overline{\W} \right)_{k,j}
\end{equation*}
We express matrix $\Q= \mathbb{E}(\Q) + \overset{\circ}{\Q}$, and obtain that
\begin{eqnarray*}
\mathbb{E} \left(\W^* \Q\W \right)_{j,k} = \frac{\sigma^2}{c_N} \left(\mathcal{T}^{(N)}_{M-L+1,L}(\mathbb{E}(\Q))\right)_{k,j} \\
- \frac{\sigma^2}{c_N} \left(\mathcal{T}^{(N)}_{M-L+1,L} (\mathbb{E}(\Q)) \mathbb{E} (\W^T \Q^T \overline{\W}) \right)_{k,j} \\
- \frac{\sigma^2}{c_N} \mathbb{E}\left(\mathcal{T}^{(N)}_{M-L+1,L}(\overset{\circ}{\Q})  \W^T \Q^T \overline{\W} \right)_{k,j}
\end{eqnarray*}
Noticing the equation,
$$
\W^T \Q^T\overline{\W}= \tilde{\Q}^T \W^T \overline{\W}
$$
we obtain that
\begin{equation*}
\mathbb{E} \left(\W^* \Q \W \right) = \frac{\sigma^2}{c_N} \mathcal{T}^{(N)}_{M-L+1,L}\left(\mathbb{E}(\Q^T)\right)
-  \frac{\sigma^2}{c_N} \mathbb{E} \left(\W^* \W \tilde{\Q}\right) \mathcal{T}^{(N)}_{M-L+1,L} \left(\mathbb{E}(\Q^T)\right)
-  \frac{\sigma^2}{c_N} \mathbb{E} \left(\W^* \W \tilde{\Q} \mathcal{T}^{(N)}_{M-L+1,L} (\overset{\circ}{\Q})\right)
\end{equation*}

Moreover we notice that
\begin{equation} \label{eq:relation-W*QW}
\W^* \Q \W = \tilde{\Q}\W^* \W = \W^* \W \tilde{\Q} = \I + z\tilde{\Q}
\end{equation}

Therefore, it holds that
\begin{equation*}
\I + z\mathbb{E}(\tilde{\Q}) =  \frac{\sigma^2}{c_N} \mathcal{T}^{(N)}_{M-L+1,L} (\mathbb{E}(\Q^T)) - \frac{\sigma^2}{c_N} (\I + z\mathbb{E}(\tilde{\Q})) \mathcal{T}^{(N)}_{M-L+1,L} \left(\mathbb{E}(\Q^T) \right) + \tilde{\Delta}
\end{equation*}
where
\begin{equation} \label{eq:Delta-tilde}
\tilde{\Delta} = - \frac{\sigma^2}{c_N} \mathbb{E} \left(\W^* \W \tilde{\Q} \mathcal{T}^{(N)}_{M-L+1,L} (\overset{\circ}{\Q})\right)
\end{equation}
This leads to the equation
\begin{equation}
\label{eq:E(tildeQ)-intermediaire}
z\mathbb{E}(\tilde{\Q}) \left( \I + \frac{\sigma^2}{c_N} \mathcal{T}^{(N)}_{M-L+1,L} (\mathbb{E}(\Q^T)) \right) = -\I + \tilde{\Delta}
\end{equation}
Lemma 2 of \cite{loubaton-bloc-hankel} (used when $(M,N)$ is replaced by $(M-L+1,N)$) implies that matrix
$$
\I + \frac{\sigma^2}{c_N} \mathcal{T}^{(N)}_{M-L+1,L} (\mathbb{E}(\Q))
$$
is invertible for $z \in \mathbb{C}^+$, and that its inverse, denoted ${\bf H}$, verifies
\begin{equation}
\label{eq:inverse-tildeH}
\left\| {\bf H} \right\|
\leq \frac{|z|}{\mathrm{Im}(z)}
\end{equation}
for $z \in \mathbb{C}^+$. (\ref{eq:E(tildeQ)-intermediaire}) implies that
$$
\mathbb{E}(\tilde{\Q}) = -\frac{\H^T}{z} + \tilde{\Delta} \H^T
$$
Therefore, (\ref{eq:convergence-esperance-forme-quadratique-tildeQW}) is equivalent to
$$
{\bf a}_N^{*} \left( -\frac{\H^T}{z} - \tilde{t}_{*}(z) {\bf I} + \tilde{\Delta} \H^T \right) {\bf b}_N \rightarrow 0
$$
Using the same technics as in Proposition 8 (see Eq. 5.3) of \cite{loubaton-bloc-hankel} as well as
(\ref{eq:inverse-tildeH}), we obtain immediately that
$$
{\bf a}_N^{*} \tilde{\Delta} \H^T {\bf b}_N \rightarrow 0
$$
It thus remains to establish that
\begin{equation}
\label{eq:E(tildeQ)-it-remains}
{\bf a}_N^{*} \left( -\frac{\H^T}{z} - \tilde{t}_{*}(z) {\bf I}  \right) {\bf b}_N \rightarrow 0
\end{equation}
For this, we use the identity
$$
-\frac{\H^T}{z} - \tilde{t}_{*}(z) {\bf I} = - \H^T \left( \frac{\I}{z\tilde{t}_{*}(z)} + (\H^T)^{-1} \right) \tilde{t}_{*}(z)
$$
$t_{*}(z)$ and $\tilde{t}_{*}(z)$ satisfy the relation $\frac{-1}{z\tilde{t}_{*}(z)} = 1 + \frac{\sigma^2}{c_N} t_{*}(z)$. Hence, the right hand side of the above equation can be written as
$$
\hspace{-0.5cm}- \H^T \left( (-1-\frac{\sigma^2}{c_N} t_{*}(z))\I + \I + \frac{\sigma^2}{c_N} \mathcal{T}^{(N)}_{M-L+1,L} \left( \mathbb{E}(\Q^T) \right)  \right) \tilde{t}_{*}(z)
$$
Corollary 1 and Theorem 2 of \cite{loubaton-bloc-hankel} imply that
$$
\left\| \mathcal{T}^{(N)}_{M-L+1,L} \left(\mathbb{E}(\Q^T) - t_{*}(z)\I\right) \right\| \rightarrow 0
$$
if $z \in \mathbb{C}^{+}$. This and (\ref{eq:inverse-tildeH}) leads to
$$
{\bf a}_N^{*} \H^T  \mathcal{T}^{(N)}_{M-L+1,L} \left(\mathbb{E}(\Q^T(z)) - t_{*}(z)\I\right) {\bf b}_N\rightarrow 0
$$
and to (\ref{eq:E(tildeQ)-it-remains}). This completes the proof of (\ref{eq:convergence-esperance-forme-quadratique-tildeQW}). \\

We now establish (\ref{eq:convergence-terme-mixte}). For this, we first remark that
for each $\theta \in \mathbb{R}$, the distribution of matrix ${\bf Z}_N e^{i \theta}$ coincides with the
distribution of ${\bf Z}_N$. Therefore, it holds that
$$
\mathbb{E}\left( {\bf Q}_N(z) {\bf Z}_N e^{i \theta} \right) = \mathbb{E}\left( {\bf Q}_N(z) {\bf Z}_N  \right)
$$
which implies that $\mathbb{E}\left( {\bf Q}_N(z) {\bf Z}_N  \right) = 0$. In order to
complete the proof of (\ref{eq:convergence-terme-mixte}), it is sufficient to establish that
if we denote by $\kappa_N$ the random variable $\kappa_N = {\bf a}_N^{*} \left( {\bf Q}_N(z) {\bf Z}_N \right) {\bf b}_N$, then, for each $p \geq 1$, it holds that
\begin{equation}
\label{eq:nash-poincare-terme-mixte}
\mathbb{E} \left| \kappa_N - \mathbb{E}(\kappa_N) \right|^{2p} = \mathcal{O}\left( \left(\frac{L}{M} \right)^{p} \right)
\end{equation}
Choosing $p$ large enough leads to $\kappa_N - \mathbb{E}(\kappa_N) = \kappa_N \rightarrow 0 \; a.s.$
as expected. (\ref{eq:nash-poincare-terme-mixte}) can be proved as above by using the
Poincar\'e-Nash inequality.

We finally justify that for each $\epsilon > 0$, (\ref{eq:convergence-forme-quadratique-Q}, \ref{eq:convergence-forme-quadratique-tildeQ}, \ref{eq:convergence-terme-mixte}) hold uniformly w.r.t. $z$ on each compact subset of $\mathbb{C} - [0, x_*^{+} + \epsilon]$. We just prove that it the case
for (\ref{eq:convergence-terme-mixte}). By item (ii), almost surely, function
$z \rightarrow \kappa_N(z)$
is analytic on $\mathbb{C} - [0, x_*^{+} + \epsilon]$. We use a standard
argument based on Montel's theorem (\cite{rudin}, p.282). We first justify that for each compact
subset $\mathcal{K} \subset \mathbb{C} - [0, x_*^{+} + \epsilon]$, then it exists a constant
$\eta$ such that
\begin{equation}
\label{eq:famille-normale}
\sup_{z \in \mathcal{K}} |\kappa_N(z)| \leq \eta
\end{equation}
for each $N$ large enough. We consider the singular value decomposition of matrix ${\bf Z}_N$:
$$
{\bf Z}_N = {\bs \Gamma}_N {\bs \Delta}_N {\bs \Theta}_N^{*}
$$
where ${\bs \Delta}_N $ represents the diagonal matrix of non zero singular values of
${\bf Z}_N$. $\kappa_N(z)$ can be written as
$$
\kappa_N(z) = {\bf a}_N^{*}  {\bs \Gamma}_N \left( {\bs \Delta}_N^{2} - z {\bf I} \right)^{-1}
{\bs \Delta}_N {\bs \Theta}_N^{*} {\bf b}_N
$$
Therefore, it holds that
$$
|\kappa_N(z)| \leq \left\| \left( {\bs \Delta}_N^{2} - z {\bf I} \right)^{-1}
{\bs \Delta}_N \right\| \| {\bf a}_N \|  \| {\bf b}_N \|
$$
Item (ii) implies that the entries of ${\bs \Delta}_N^{2}$ are located into $[0, x_*^{+} + \epsilon]$.
for each $N$ large enough. Therefore, for each $z \in \mathcal{K}$, it holds that
$$
\left\| \left( {\bs \Delta}_N^{2} - z {\bf I} \right)^{-1}
{\bs \Delta}_N \right\| \leq \frac{1}{\mathrm{dist}([0, x_*^{+} + \epsilon], \mathcal{K})}
$$
The conclusion follows from the hypothesis that vectors ${\bf a}_N$ and ${\bf b}_N$
satisfy $\sup_{N} (\|{\bf a}_N\|, \|{\bf b}_N\|) < +\infty$. (\ref{eq:famille-normale}) implies that the sequence of analytic
functions $(\kappa_N)_{N \geq 1}$ is a normal family . Therefore, it exists a subsequence extracted from $(\kappa_N)_{N \geq 1}$ that converges uniformly  on each compact subset of $\mathbb{C} - [0, x_*^{+} + \epsilon]$ towards a certain analytic function $\kappa_*$. As (\ref{eq:convergence-terme-mixte})
holds for each $z \in \mathbb{C}^{+}$, function $\kappa_*$ is identically zero. We have thus shown that
each converging subsequence extracted from $(\kappa_N)_{N \geq 1}$ converges uniformly towards
$0$ on each compact subset of $\mathbb{C} - [0, x_*^{+} + \epsilon]$. This, in turn, shows that
the whole sequence converges uniformly on each compact subset of $\mathbb{C} - [0, x_*^{+} + \epsilon]$
as expected.

\bibliographystyle{plain}

\end{document}